\RequirePackage{fix-cm}
\documentclass[twocolumn,epjc3]{svjour3}  
\smartqed  
\RequirePackage{graphicx}
\RequirePackage{latexsym}
\RequirePackage[numbers,sort&compress]{natbib}
\RequirePackage[colorlinks,citecolor=blue,urlcolor=blue,linkcolor=blue]{hyperref}
\usepackage{amsmath}
\usepackage{amssymb}
\usepackage{widetext}
\usepackage{subfigure}

\journalname{Eur. Phys. J. C}
\begin{document}

\title{Inflationary quantum dynamics and backreaction using a classical-quantum correspondence
}


\author{Reginald Christian Bernardo\thanksref{e1}
}

\thankstext{e1}{e-mail: rbernardo@nip.upd.edu.ph}


\institute{
National Institute of Physics, University of the Philippines, Diliman, Quezon City 1101, Philippines
}

\date{Received: date / Accepted: date}

\maketitle

\begin{abstract}
We study inflationary dynamics using a recently introduced classical-quantum correspondence for investigating the  backreaction of a quantum mechanical degree of freedom to a classical background. Using specifically a coupled Einstein-Klein-Gordon system, an approximation that holds well during the very early inflationary era when modes are very deep inside the Hubble horizon, we show that the backreaction of a mode of the quantum field will renormalize the Hubble parameter only if the mode's wavelength is longer than some threshold Planckian length scale. Otherwise, the mode will destabilize the inflationary era. We also present an approximate analytical solution that supports the existence of such short-wavelength threshold and compare the results of the classical-quantum correspondence with the traditional perturbative-iterative method in semiclassical gravity.
\end{abstract}


\section{Introduction}
\label{sec:introduction}

The existence of an inflationary era that precedes the standard radiation-(nonrelativistic) matter-dark energy expansion history is a powerful hypothesis that simultaneously solves the horizon, flatness, and monopole problems in cosmology \cite{STAROBINSKY198099, inflation_guth, inflation_linde, inflation_albrecht}. Most importantly, inflation brings into the overall cosmological picture a quantum mechanical factory of inhomogeneities in the early Universe that could later on seed structure formation \cite{Starobinsky:1979ty, Mukhanov:1981xt, HAWKING1982295, Starobinsky:1982ee, inflation_guth_2, inflation_reheating_albrecht, Starobinsky:1983zz, structure_formation_garcia}. Observational signatures of primordial inflation can be found in the cosmic microwave background and possibly in the future through gravitational wave observations \cite{constraint_davis_1992, constraint_bartolo_2004, constraint_bartolo_2016, constraint_kamionkowski_2016, constraint_kuroyanagi_2018, constraint_antusch_2017, inflation_chowdury_2019, constraint_shokri_2019, constraint_brandenberger_2019}. 

The analysis of inflationary particle production is typically done within the confines of semiclassical gravity, wherein the modes, or rather particles, of a quantum field are drawn out of the vacuum by the expanding spacetime. However, the eventual response of the background spacetime to an accumulated cosmic fluid -- the \textit{backreaction} -- is mostly restricted,  if not completely overlooked, because of understandable technical reasons. Simply put, at the linear order formalism, the classical background merely acts as a fixed stage on which the perturbations behave as test fields. The backreaction effect causing the stage to respond to the fields can therefore be found at the higher order, nonlinear interactions. Several breakthroughs have been made even with the existing challenges and the analysis of quantum backreaction during inflation can be expected to remain as an active field in forthcoming years \cite{Starobinsky:1986fx, Starobinsky:1994bd, inflation_backreaction_mukhanov, inflation_backreaction_tsamis, inflation_backreaction_tsamis_2, Finelli:2008zg, Finelli:2010sh, Vennin:2015hra, inflation_backreaction_brandenberger_2015, inflation_backreaction_miao_2017, inflation_backreaction_brandenberger_2018, inflation_backreaction_bloomfield, inflation_preheating_armendariz}. Inflationary particle production and backreaction must be cautiously dealt with all the same and any method for shedding light to this puzzle even in simpler analog-physical systems should be worth investigating.

A classical-quantum correspondence (CQC) was recently introduced to study the backreaction of a quantum harmonic oscillator (QHO) to a classical background \cite{cqc_backreaction}. Indeed, the dynamics of a quantum mechanical degree of freedom (d.o.f.) could be treated simply by using the classical Bogoliubov coefficients, leading to Hawking radiation and inflationary particle production. The CQC, on the other hand, shows that a complexified classical harmonic oscillator (CHO) could be used, instead of the Bogoliubov coefficients, to look at the backreaction of a QHO to a classical background. This complexified oscillator retains the quantum mechanical d.o.f.s just as much as the Bogoliubov coefficients but it is arguably simpler, and more pedagogical, because it can be treated using integration methods familiar for dealing with the CHO. We clarify that the above methodology for dealing with the backreaction, i.e., utilizing a complexified CHO to deal with the coupled QHO-background dynamics, is what is meant by classical-quantum correspondence throughout this paper. In particular, we emphasize that this is different from perhaps synonymous terms pertaining to the important transition of quantum excitations to classical modes due to an expanding classical background \cite{Albrecht:1992kf, Polarski:1995jg, Kiefer:2008ku}. Physical insights on backreaction have already been drawn by using the CQC in its early applications to a rolling ball-QHO system \cite{cqc_backreaction} and Hawking radiation \cite{cqc_hawking_radiation} but it is the elegance by which these important results have been obtained that speaks loudly about the potential of the method. Moreover, the CQC has been extended to cover field theory applications \cite{cqc_fields} and applied for instance to a coupled scalar field system \cite{cqc_rolling_scalar}, one of which is the rolling inflaton familiar in the inflationary context, and the evaporation of breathers \cite{cqc_breathers}. It is important to stress out that the CQC is an exact solution to the field equations and that perhaps its only limitation is whenever the coupled classical-quantum d.o.f.s picture no longer holds. Applying the CQC to more systems in order to obtain physical insights on quantum mechanical backreaction and to assess CQC's regime of validity is therefore of considerable importance. In this work, we apply the CQC for the first time in a cosmological background, specifically on inflationary dynamics.

We consider a coupled Einstein-(massless) Klein-Gordon system (Eqs. \eqref{eq:wave_equation} and \eqref{eq:einstein_equation}), wherein the scalar field is quantized and driven by the cosmological dynamics. This model is admitedly not a faithful representation of the entire history of the early Universe but it holds well during the very early stages of inflation when the modes are deep inside the Hubble horizon \cite{weinberg}. At the same time, it appears that this model is one that is tailored for an exhibition of the CQC in the inflationary setting. Two specific reasons make it particularly appealing. First, it is the simplest and most natural testbed for analyzing the semiclassical effects of gravity. It is also quite flexible in the sense that the case of a massive scalar field can be  accommodated by replacing the wavenumber $k$ by $\sqrt{ k^2 + m^2}$ where $m$ is the bare mass of the quantum field. Second, the d.o.f. of a massless scalar field can be used to describe the d.o.f.s of higher spin fields, e.g., the two independent polarizations of the massless tensor field. The results obtained in this paper could therefore be used to also gain insight on the backreaction of the gravitational waves produced during inflation. The same model, however, cannot be used to describe the production of fermions. This calls for an independent work using a coupled Einstein-Maxwell-Dirac system.

The main objective of this work is to demonstrate the relative ease of implementing the CQC and to gain insight on backreaction in an inflationary setting. The rest of this work proceeds as follows. We start by revisiting the CQC in detail and its application to the rolling ball-QHO system (Section \ref{sec:cqc}). This prepares the application to inflationary particle production (Section \ref{sec:backreaction_ds}). We recast the field equations of the coupled Einstein-Klein-Gordon system into a form that can be objectively studied using the CQC (Section \ref{subsec:einstein_kg_system}) and present the results of various numerical integrations (Section \ref{subsec:numerical_investigation}). We furthermore compare the CQC to the perturbative-iterative method (Section \ref{sec:cqc_iterative_method}) and discuss the relevant insights and inherent limitations of the present work (Section \ref{sec:discussion}).

We work with the mostly-plus metric signature $( - + + + )$ and Planckian units $c = 8 \pi G = \hbar = 1$ where $c$, $G$, and $\hbar$ are the speed of light in vacuum, Newton's gravitational constant, and Planck's constant, respectively (\ref{sec:planck_units}). 

\section{Classical-quantum correspondence}
\label{sec:cqc}

In this section we provide a detailed overview of the classical-quantum correspondence (CQC) and revisit its application to a toy problem. This prepares the application to the cosmological background.

\subsection{An overview of the CQC}

The Hamiltonian for a driven QHO with time-dependent frequency is given by 
\begin{equation}
\hat{H} = \frac{\hat{p}^2}{2} + \frac{ \omega \left( t \right)^2}{2}  \hat{x}^2
\end{equation}
where $\omega \left( t \right)$ is the \textit{classical} driving frequency or the \textit{background}. As is well-known, this Hamiltonian can be diagonalized in the Heisenberg picture by writing it as 
\begin{equation}
\hat{H} = \omega \left( t \right) \left( \hat{a}^\dagger \hat{a} + \frac{1}{2} \right)
\end{equation}
where $\hat{a}$ and $\hat{a}^\dagger$ are the annihilation and creation operators, respectively, i.e., $\left[ a , a^\dagger \right] = 1$, given by
\begin{eqnarray}
\hat{a} &=& \frac{ \hat{p} - i  \omega \hat{x} }{ \sqrt{2 \omega} } \\
\hat{a}^\dagger &=& \frac{ \hat{p} + i  \omega \hat{x} }{ \sqrt{2 \omega} } .
\end{eqnarray}
Solving operator equations of motion, however, is often extremely challenging \footnote{
The dynamics of an operator $\hat{A}$ is given by $i \frac{d \hat{A}}{dt} = \left[ \hat{A}, \hat{H} \right] + i \frac{\partial \hat{A}}{\partial t}$ where $\hat{H}$ is the system's Hamiltonian.
}. For this reason, the quantum dynamical problem becomes more manageable after using the Bogoliubov transformation
\begin{eqnarray}
\hat{a} \left( t \right) &=& \alpha \left( t \right) \hat{a}_0 + \beta \left( t \right) \hat{a}^\dagger_0 \\
\hat{a}^\dagger \left( t \right) &=& \alpha^* \left( t \right) \hat{a}^\dagger_0 + \beta^* \left( t \right) \hat{a}_0
\end{eqnarray}
where $\alpha$ and $\beta$ are classical variables and $\hat{a}_0 = \hat{a} \left( 0 \right)$ and $\hat{a}^\dagger_0 = \hat{a}^\dagger \left( 0 \right)$ are the initial values of the annihilation and creation operators. The Heisenberg equations of motion for the operators $\hat{a}$ and $\hat{a}^\dagger$ become
\begin{eqnarray}
\dot{\alpha} &=& -i\omega \alpha	- \frac{\dot{\omega}}{2\omega} \beta^* \\
\dot{\beta} &=& -i\omega \beta - \frac{\dot{\omega}}{2\omega} \alpha^*
\end{eqnarray}
and it can be shown that the Bogoliubov coefficients $\alpha$ and $\beta$ satisfy
\begin{eqnarray}
|\alpha|^2 - |\beta|^2 &=& 1 \\
\beta \dot{\alpha} - \alpha \dot{\beta} &=& \frac{\dot{\omega}}{2\omega} .
\end{eqnarray}
Furthermore, in terms of the Bogoliubov coefficients, the vacuum expectation value of the Hamiltonian can be written as
\begin{equation}
\langle 0 | \hat{H} | 0 \rangle = \omega \left( t \right) \left( |\beta|^2 + \frac{1}{2} \right)
\end{equation}
where $| 0 \rangle$ is the vacuum state at $t = 0$, i.e., $\hat{a}_0 | 0 \rangle = 0$. This quantifies the excitation driven by the background $\omega \left( t \right)$.

Quite remarkably, it was shown in Ref. \cite{cqc_backreaction} that an improvement, and arguably a simplification, to the traditional Bogoliubov transformation method can be further obtained by using the following transformation 
\begin{eqnarray}
\alpha &=& \frac{1}{ \sqrt{ 2\omega } } \left( \dot{z}^* - i \omega z^* \right) \\
\beta &=& \frac{1}{ \sqrt{ 2\omega } } \left( \dot{z} - i \omega z \right) \\
z &=& \xi + i \chi 
\end{eqnarray}
which trades the Bogoliubov coefficients $\alpha$ and $\beta$ for a single complex variable $z$. This variable $z$ describes a complex classical harmonic oscillator (CHO),
\begin{equation}
\ddot{z} + \omega^2 \left( t \right)z = 0 ,
\end{equation}
and in terms of which the original QHO problem can be shown to be equivalent to that of two CHOs with constrained initial conditions. Indeed, it can be shown that the real and imaginary parts of $z$ independently satisfy the CHO equation of motion,
\begin{eqnarray}
\ddot{\xi} + \omega^2 \left( t \right) \xi &=& 0 \\
\ddot{\chi} + \omega^2 \left( t \right) \chi &=& 0 ,
\end{eqnarray}
but with the constrained initial conditions (coming from $\alpha \left( 0 \right) = 1$ and $\beta \left( 0 \right) = 0$) 
\begin{eqnarray}
\xi_0 &=& 0 \\
\dot{\xi}_0 &=& \sqrt{\omega_0/2} \\
\chi_0 &=& -1/\sqrt{2 \omega_0} \\
\dot{\chi}_0 &=& 0 \\
\xi \dot{\chi} - \chi \dot{\xi} &=& 1/2
\end{eqnarray}
where $\omega_0 = \omega \left( 0 \right)$. Obviously, $\xi$ and $\chi$ are also CHOs where $\xi$ is initially pushed to the right away from the origin and $\chi$ is initially maximally stretched to the left and released from rest. In other words, the CHOs $\xi$ and $\chi$ are out of phase by $\pi / 2$. 
Furthermore, in terms of the CHOs, the vacuum expectation value can be written simply as the sum of the energies of two driven CHOs, i.e., 
\begin{equation}
\langle 0 | \hat{H} | 0 \rangle = E_\xi + E_\chi
\end{equation}
where $E_\xi$ and $E_\chi$ are given by
\begin{eqnarray}
E_\xi &=& \frac{ \dot{\xi}^2 }{2} + \frac{\omega^2 \xi^2}{2} \\
E_\chi &=& \frac{ \dot{\chi}^2 }{2} + \frac{ \omega^2 \chi^2 }{2} .
\end{eqnarray}
Shortly, we shall revisit the application of the above transformation to the problem of a QHO coupled to a classical rolling ball (Section \ref{subsec:toy_problem}). 

It is possible work out the direct transformation from the original QHO operators $\hat{x}$ and $\hat{p}$ to the CHOs $\xi$ and $\chi$ \footnote{
To see this, one should simply track the transformations used throughout. For instance, this procedure leads to $\hat{x} = z^* \hat{a}_0 + z \hat{a}_0^\dagger$ and $\hat{p} = m \left(  \dot{z}^* \hat{a}_0 + \dot{z} \hat{a}_0^\dagger \right)$ which leads straight to Eqs. \eqref{eq:x_cqc} and \eqref{eq:p_cqc}.
}. The resulting transformation is given by 
\begin{eqnarray}
\label{eq:x_cqc} \hat{x} &=& \sqrt{ \frac{2}{ \omega_0} } \xi \hat{p}_0 - \sqrt{2 \omega_0} \chi \hat{x}_0 \\
\label{eq:p_cqc} \hat{p} &=& \sqrt{ \frac{2}{\omega_0} } \dot{\xi} \hat{p}_0 - \sqrt{2 \omega_0} \dot{\chi} \hat{x}_0 
\end{eqnarray}
where $\hat{x}_0$ and $\hat{p}_0$ are the initial values of the position and momentum operators, respectively, at $t = 0$.

\subsection{Application to a toy problem: classical rolling particle and QHO}
\label{subsec:toy_problem}

Consider the toy problem described by the Hamiltonian \cite{cqc_backreaction} 
\begin{equation}
\label{eq:hamiltonian_toy}
\hat{H} = \frac{p_y^2}{2} - a y + \frac{\hat{p}_x^2}{2} + \frac{\omega_0^2}{2}  \hat{x}^2 + \frac{\lambda}{2} y^2 \hat{x}^2 .
\end{equation}
In this problem, the QHO d.o.f. is given by $\hat{x}$ and the classical rolling particle d.o.f. is given by $y$. Also, $\omega_0$ and $a$ are constants that would be the QHOs natural frequency and the rolling ball's acceleration in the uncouplied limit. It should be stressed out that the classical terms in Eq. \eqref{eq:hamiltonian_toy}, e.g., $p_y^2 / 2$, should always be viewed besides the identity operator $\hat{I}$ for consistency with the explicitly operator terms. 

To proceed using the CQC, we write down the Hamiltonian in terms of CHOs described by Eqs. \eqref{eq:x_cqc} and \eqref{eq:p_cqc}. This straightforward calculation leads to
\begin{equation}
\label{eq:hamiltonian_toy_cqc}
\begin{split}
\hat{H} = \frac{p_y^2}{2} & - a y 
+ \frac{1}{2} \bigg( \frac{2}{\omega_0} \dot{\xi}^2 \hat{p}_0^2 + 2 \omega_0 \dot{\chi}^2 \hat{x}_0^2 \\
& \phantom{gggggggggggggggg} - 2 \dot{\xi} \dot{\chi} \left( \hat{p}_0 \hat{x}_0 + \hat{x}_0 \hat{p}_0 \right)  \bigg) \\
& + \frac{1}{2} \left( \omega_0^2 + y^2 \right) \bigg( \frac{2}{\omega_0} \xi^2 \hat{p}_0^2 + 2 \omega_0 \chi^2 \hat{x}_0^2\\
& \phantom{gggggggggggggggg} - 2 \xi \chi \left( \hat{p}_0 \hat{x}_0 + \hat{x}_0 \hat{p}_0 \right) \bigg) .
\end{split}
\end{equation}
The classical equation of motion for $y$ can be derived from this simply first by taking the vacuum expectation value and using the Hamiltonian equations,
\begin{eqnarray}
\dot{y} &=& \partial_{p_y} \langle 0 | \hat{H} | 0 \rangle \\
\dot{p}_y &=& - \partial_y \langle 0 | \hat{H} | 0 \rangle  .
\end{eqnarray}
This leads to 
\begin{equation}
\label{eq:eom_y_toy}
\begin{split}
\ddot{y} = a - \lambda y \bigg( & \frac{2}{\omega_0} \xi^2 \langle 0 | \hat{p}_0^2 | 0 \rangle + 2 \omega_0 \chi^2 \langle 0 | \hat{x}_0^2 | 0 \rangle \\
& \phantom{gggggggg} - 2 \xi \chi \langle 0 | \left( \hat{p}_0 \hat{x}_0 + \hat{x}_0 \hat{p}_0 \right) | 0 \rangle \bigg) .
\end{split}
\end{equation}
Clearly, this shows that the rolling ball would be uninterupted and fall with constant acceleration $a$ provided that there's no coupling with the QHO $(\lambda = 0)$. The backreaction of the QHO to the background $y$ is therefore quantified by the terms attached to $\lambda$ in Eq. \eqref{eq:eom_y_toy}. Before this can be used, however, it is important to specify the initial shape of the wave packet as
\begin{equation}
\label{eq:gaussian_wave_packet_ho}
\Psi \left( t = 0 \right) = \left( \frac{\omega_0}{\pi \hbar } \right)^{1/4} e^{- \omega_0 x^2 / 2 }
\end{equation}
as this is a quantum dynamics problem \footnote{
It is natural to choose Eq. \eqref{eq:gaussian_wave_packet_ho} as it is the ground state wave function of the time-independent QHO. However, other choices are possible. We leave the investigation of the dependence of the dynamics on the choice of initial condition for future work. 
}. Using Eq. \eqref{eq:gaussian_wave_packet_ho}, it can be shown that
\begin{eqnarray}
\langle 0 | \hat{p}_0^2 | 0 \rangle &=& \omega_0/2 \\
\langle 0 | \hat{x}_0^2 | 0 \rangle &=& 1 / \left( 2 \omega_0 \right) \\
\langle 0 | \hat{p}_0 \hat{x}_0 + \hat{x}_0 \hat{p}_0 | 0 \rangle &=& 0 
\end{eqnarray}
and Eq. \eqref{eq:eom_y_toy} reduces to
\begin{equation}
\label{eq:eom_y_toy_reduced}
\ddot{y} = a - \lambda y \left( \xi^2 + \chi^2 \right) .
\end{equation}
The complete dynamics incorporating the backreaction of the QHO to the background can finally be obtained by solving the coupled classical system given by Eq. \eqref{eq:eom_y_toy_reduced} and the CHO equations of motion
\begin{eqnarray}
\ddot{\xi} &=& - \left( \omega_0^2 + \lambda y^2 \right) \xi \\
\ddot{\chi} &=& - \left( \omega_0^2 + \lambda y^2 \right) \chi 
\end{eqnarray}
together with the initial conditions
\begin{eqnarray}
y_0 &=& 0 \\
\dot{y}_0 &=& 0 \\
\xi_0 &=& 0 \\
\dot{\xi}_0 &=& \sqrt{\omega_0 / 2} \\
\chi_0 &=& -1/\sqrt{2 \omega_0} \\
\dot{\chi}_0 &=& 0 .
\end{eqnarray}
The vacuum expectation value could also be simply calculated by adding the energies of the rolling ball and the CHOs, i.e.,  
\begin{equation}
\langle 0 | \hat{H} | 0 \rangle = E_y +  E_\xi + E_\chi .
\end{equation}
In Ref. \cite{cqc_backreaction} it was shown using this system of equations that backreaction renormalizes the bare acceleration $a$ of the background.

\section{Backreaction of quantum scalar field on a cosmological background}
\label{sec:backreaction_ds}

In this section, we use the CQC to study the evolution of a mode in the cosmological background,
\begin{equation}
\label{eq:frw}
ds^2 = - dt^2 + a^2 \left( t \right) d \vec{x}^2 ,
\end{equation}
where $a$ is the scale factor. We setup the field equations in Section \ref{subsec:einstein_kg_system} and present the results in Section \ref{subsec:numerical_investigation}. 

\subsection{Setup: Einstein-Klein-Gordon system}
\label{subsec:einstein_kg_system}

We rely on the coupled Einstein-Klein-Gordon theory. In this way, the quantum field $\hat{\phi}$ satisfies the (massless) Klein-Gordon equation
\begin{equation}
\label{eq:wave_equation}
\Box \hat{\phi} = 0
\end{equation}
and the metric (the background) satisfies the Einstein equation
\begin{equation}
\label{eq:einstein_equation}
G_{\alpha \beta} = 8 \pi \left( T^{(M)}_{\alpha \beta} + \langle 0 | \hat{T}^{(\phi)}_{\alpha \beta} | 0 \rangle \right) 
\end{equation}
where variables with hats, e.g., $\hat{\phi}$, are (quantum mechanical) operators.  
Also, in Eq. \eqref{eq:einstein_equation}, the quantity $\langle 0 | \hat{T}^{(\phi)}_{\alpha \beta} | 0 \rangle$ is the vacuum expectation value of the scalar field's stress-energy tensor (SET), $\hat{T}^{(\phi)} \sim \hat{\phi}^2, \left( \partial \hat{\phi} \right)^2$, and $T^{(M)}_{\alpha \beta}$ is the matter sector's SET. To support the inflationary era, we assume that the matter sector is completely dominated by vacuum energy, a barotropic perfect fluid with an equation of state $w = P/\rho = -1$, where $\rho$ and $P$ are the energy density and pressure, respectively \footnote{
A more rigorous way that is compatible with inflationary theories is to consider a slowly rolling inflaton, e.g., $\hat{\Psi}$, in the place of vacuum energy. We leave this for future work.
}.
On the other hand, the SET of the quantum scalar field is given by
\begin{equation}
\label{eq:stress_energy_scalar}
\hat{T}^{(\phi)}_{\alpha \beta} = \left( \partial_\alpha \hat{ \phi } \right) \left( \partial_\beta \hat{ \phi } \right) - \frac{1}{2} g_{\alpha \beta} \left( \partial \hat{\phi} \right)^2
\end{equation}
which could be identified as a fluid with an energy density $\hat{ \rho }^{(\phi)}$ and pressure $\hat{P}^{(\phi)}$ given by
\begin{equation}
\hat{P}^{(\phi)} = \hat{ \rho }^{(\phi)} = - \frac{1}{2} \left( \partial \hat{ \phi } \right)^2 .
\end{equation}
This identification also shows that particles driven out of the vacuum by the expansion could be described by an effective equation of state, $w_\phi = \langle \hat{P}^{(\phi)} \rangle / \langle \hat{ \rho }^{(\phi)} \rangle = 1$, where the averaging, $\langle \cdots \rangle$, is taken over a quantum state. Thus, because cosmic fluids with an equation of state $w > -1/3$ contribute to deceleration, it is reasonable to expect the production of $\hat{\phi}$-particles and their backreaction to the spacetime would slow down the ongoing expansion. 

An important comment is in order. First of all, the object $\langle 0 | \hat{T}^{(\phi)}_{\alpha \beta} | 0 \rangle$ sourcing the response of the background (encoded in the Einstein tensor $G_{\alpha \beta}$) in Eq. \eqref{eq:einstein_equation} is a generally divergent quantity due to the sum of albeit finite but infinitely many zero point energies coming from each of the modes of $\hat{\phi}(x)$. This divergence can be dealt with using renormalization and has been the topic of many important papers in the field \cite{Zeldovich:1971mw, Fulling:1974pu, Mamaev:1976zb}. On the other hand, we shall deal with only a mode of the quantum field in what follows to obtain finite predictions of the backreaction per mode. This goes well by the scope of the paper which is to demonstrate the application of the CQC for the first time in an expanding spacetime background. However, the renormalization of the zero point fluctuations within the context of backreaction and the CQC will have to be confronted in a future work.

We proceed to setting up the dynamical system. To achieve this our first goal is to rewrite Eqs. \eqref{eq:wave_equation} and \eqref{eq:einstein_equation} into a QHO driven by an expanding spacetime. In the cosmological background (Eq. \eqref{eq:frw}), the Klein-Gordon equation reduces to
\begin{equation}
\label{eq:kg_ds}
\ddot{ \hat{\phi} } + 3 H \dot{ \hat{\phi} } - \frac{ \nabla^2 \hat{\phi} }{a^2} = 0 
\end{equation}
and the Einstein equation (Eq. \eqref{eq:einstein_equation}) turns into the Friedmann equations
\begin{eqnarray}
3 H^2 &=& \rho + \langle 0 | \hat{ \rho }^{(\phi)} | 0 \rangle \\
2 \dot{ H } + 3 H^2 &=& - P - \langle 0 | \hat{ P }^{(\phi)} | 0 \rangle ,
\end{eqnarray}
where $H$ is the Hubble parameter, $H = \dot{ a } / a$, and an overdot indicates differentiation with respect to the comoving time $t$. Recognizing that the Klein-Gordon equation (Eq. \eqref{eq:kg_ds}) must be restructured as a driven QHO problem to use the CQC, we deliberately express the field equations in terms of the conformal time $\eta$ and conformal scalar field $\hat{\psi}$ defined by \footnote{
In terms of the conformal field $\hat{\psi} = a \hat{\phi}$ and the comoving time $t$, the Klein-Gordon equation is given by
\begin{equation}
- \ddot{ \hat{\psi} } - H \dot{ \hat{\psi} } + \left( H^2 + \frac{ \ddot{a} }{a} \right) \hat{ \psi } + \frac{ \nabla^2 \hat{\psi} }{a^2} = 0 . 
\end{equation}
On the other hand, the transformation from comoving time $t$ to conformal time $\eta$ is given by
\begin{eqnarray}
\ddot{f} &=& \frac{f''}{a^2} - \frac{H f'}{a} \\
\dot{f} &=& \frac{f'}{a}
\end{eqnarray}
where $H = \dot{a}/a = a'/a^2$.
}
\begin{equation}
\dot{\eta} = 1/a 
\end{equation}
and 
\begin{equation}
\hat{ \psi } = a \hat{ \phi } ,
\end{equation}
respectively. This leads to
\begin{equation}
\label{eq:kg_conf}
\hat{ \psi }''  - \left( \nabla^2 + \frac{ a'' }{a} \right) \hat{ \psi } = 0 
\end{equation}
where a prime denotes differentiation with respect to the conformal time $\eta$. The above equation can easily be recognized as the QHO's equation of motion for the field $\hat{\psi}$ at a given location $\vec{x} = \vec{X}$.
We must also express the Friedmann equations in terms of the conformal variables $\eta$ and $\hat{\psi}$. Straightforward calculation leads to
\begin{eqnarray}
\label{eq:fe1_conf} 3 \frac{a^{\prime 2}}{a^4} &=& \rho + \langle 0 | \hat{ \rho }^{(\psi)} | 0 \rangle \\
\label{eq:fe2_conf} 2 \frac{a''}{a^3} - \frac{a^{\prime 2}}{a^4} &=& - P - \langle 0 | \hat{ P }^{(\psi)} | 0 \rangle
\end{eqnarray}
where the energy density and pressure of the scalar field become
\begin{equation}
\label{eq:rho_psi_general}
\begin{split}
 \hat{ \rho }^{(\psi)} =  \hat{ P }^{(\psi)} = \frac{1}{2 a^4} \bigg[ \hat{\psi}^{\prime 2} & - \frac{a'}{a} \left( \hat{\psi} \hat{\psi}' + \hat{\psi}' \hat{\psi} \right) \\
& + \frac{a^{\prime 2}}{a^2} \hat{\psi}^2 - \left( \nabla \hat{\psi} \right)^2 \bigg] .
\end{split}
\end{equation}
This achieves our first goal. Eqs. \eqref{eq:kg_conf}, \eqref{eq:fe1_conf}, and \eqref{eq:fe2_conf} indeed make up a system that could be solved using the CQC.

The above set of equations, however, while it could be considered already as a QHO problem coupled to a dynamical background, contains infinitely many QHOs defined at each point $\vec{x} = \vec{X}$. To simplify the analysis, we instead focus only on a single $k$-mode, $\hat{ \psi }_k \left( \eta \right)$, of the field $\hat{\psi}  \left( \eta, \vec{x} \right)$ defined by the Fourier transform 
\begin{equation}
\label{eq:fourier}
\hat{\psi} \left( \eta, \vec{x} \right) \sim e^{ i \vec{k} \cdot \vec{x} } \hat{ \psi }_k \left( \eta \right) .
\end{equation}
In this way, we work instead with modes $\hat{ \psi }_k \left( \eta \right)$ which satisfy the driven QHO equation of motion
\begin{equation}
\label{eq:kg_conformal_k}
\hat{ \psi }_k '' + \omega^2 \left( \eta \right) \hat{ \psi }_k = 0 
\end{equation}
with the time dependent frequency
\begin{equation}
\label{eq:omega_eta}
\omega^2 \left( \eta \right) = k^2 - \frac{a''}{a} . 
\end{equation}
The energy density and pressure of a mode are given by
\begin{equation}
\label{eq:rho_k_mode}
\begin{split}
 \hat{ \rho }_k^{(\psi)} =  \hat{ P }^{(\psi)}_k = \frac{1}{2 a^4} \bigg[ \hat{\psi}_k^{\prime 2} & + \left( k^2 + \frac{a^{\prime 2}}{a^2} \right) \hat{\psi}_k^2 \\
&  - \frac{a'}{a} \left( \hat{\psi}_k \hat{\psi}_k' + \hat{\psi}'_k \hat{\psi}_k \right) \bigg] .
\end{split}
\end{equation}
This sets the stage for the CQC. Noting that each mode $\hat{\psi}_k$ is described by the Lagrangian
\begin{equation}
L_k = \frac{ \psi_k^{\prime 2}}{2}  - \frac{\omega^2 \left( \eta \right)}{2} \psi_k^2  ,
\end{equation}
the Hamiltonian is therefore given by
\begin{equation}
\hat{ H }_k = \frac{ \hat{p}_k^2}{2} + \frac{1}{2} \omega^2 \left(\eta\right) \hat{ \psi }_k^2
\end{equation}
where
\begin{equation}
p_k = \frac{\partial L_k}{\partial \psi_k'} = \psi_k'
\end{equation}
is the field conjugate to $\psi_k$.

Analogous to Eqs. \eqref{eq:x_cqc} and \eqref{eq:p_cqc}, we implement the CQC by performing the transformation
\begin{eqnarray} 
\label{eq:psi_cqc} \hat{\psi}_k &=& \sqrt{ \frac{2}{\omega_0} } \xi \hat{p}_{k0} - \sqrt{2\omega_0} \chi \label{eq:cqc_p_k_mm} \hat{\psi}_{k0} \\
\label{eq:psi_conj_cqc} \hat{p}_k &=& \sqrt{ \frac{2}{\omega_0} } \xi' \hat{p}_{k0} - \sqrt{2\omega_0} \chi' \hat{\psi}_{k0}
\end{eqnarray}
where $\left[ \hat{\psi}_{k0}, \hat{p}_{k'0} \right] = i \delta_{kk'}$ and $\left( \hat{\psi}_{k0}, \hat{p}_{k0} \right)$ are the operators $\left( \hat{ \psi }_k, \hat{p}_k \right)$ at $\eta = \eta_0$. This converts the driven QHO dynamical problem (Eq. \eqref{eq:kg_conformal_k}) into that of two driven CHOs $\xi$ and $\chi$ described by
\begin{eqnarray}
\label{eq:cqc_xi} \xi'' + \omega^2 \left(\eta\right) \xi &=& 0 \\
\label{eq:cqc_chi} \chi'' + \omega^2 \left(\eta\right) \chi &=& 0
\end{eqnarray}
with the initial conditions
\begin{eqnarray}
\xi \left( \eta_0 \right) &=& 0 \\ 
\xi' \left( \eta_0 \right) &=& \sqrt{\omega_0 / 2} \\
\chi \left( \eta_0 \right) &=& - 1/ \sqrt{2 \omega_0} \\
\chi' \left( \eta_0 \right) &=& 0 
\end{eqnarray}
where $\omega_0 = \omega \left( \eta_0 \right)$. In terms of the CHOs, the vacuum expectation value of the mode Hamiltonian $\hat{H}_k$ can be written as
\begin{equation}
\langle 0 | \hat{ H }_k | 0 \rangle = E_\xi + E_\chi 
\end{equation}
where $E_\xi$ and $E_\chi$ are the total energies of the CHOs. The Friedmann equations (Eqs. \eqref{eq:fe1_conf} and \eqref{eq:fe2_conf}) and Eqs. \eqref{eq:cqc_xi} and \eqref{eq:cqc_chi} form the CQC dynamical system for the vector $\left( a, \xi, \chi \right)$ which describes cosmological dynamics incorporating single $k$-mode particle production and backreaction.

We complete the setup by expressing the vacuum expectation values $\langle 0 | \hat{ \rho }^{(\psi)} | 0 \rangle$ and $\langle 0 | \hat{ P }^{(\psi)} | 0 \rangle$ in terms of the CHOs $\left( \xi, \chi \right)$ and preparing the system in some initial state. To do so, we note that Eqs. \eqref{eq:psi_cqc} and \eqref{eq:psi_conj_cqc} straighforwardly lead to
\begin{equation}
\label{eq:psi_k_2_vac}
\begin{split}
\langle 0 | \hat{\psi}_k^2 | 0 \rangle = & \frac{2}{ \omega_0 } \xi^2 \langle 0 | \hat{p}_{k0}^2 | 0 \rangle + 2 \omega_0 \chi^2 \langle 0 | \hat{\psi}_{k0}^2 | 0 \rangle \\
& \phantom{ggggg} - 2 \xi \chi \langle 0 | \left( \hat{p}_{k0} \hat{ \psi }_{k0} + \hat{ \psi }_{k_0} \hat{p}_{k0} \right) | 0 \rangle ,
\end{split}
\end{equation}
\begin{equation}
\begin{split}
\label{eq:p_k_2_vac}
\langle 0 | \hat{p}_k^2 | 0 \rangle = & \frac{2}{ \omega_0 } \xi^{\prime 2} \langle 0 | \hat{p}_{k0}^2 | 0 \rangle + 2 \omega_0 \chi^{\prime 2} \langle 0 | \hat{\psi}_{k0}^2 | 0 \rangle \\
& \phantom{ggggg} - 2 \xi' \chi' \langle 0 | \left( \hat{p}_{k0} \hat{ \psi }_{k0} + \hat{ \psi }_{k_0} \hat{p}_{k0} \right) | 0 \rangle  ,
\end{split}
\end{equation}
and
\begin{equation}
\label{eq:psi_k_p_k_vac}
\begin{split}
\langle 0 | \left( \hat{\psi}_k \hat{p}_k + \hat{p}_k \hat{\psi}_k \right) | 0 \rangle = & \frac{4}{\omega_0} \xi \xi' \langle 0 | \hat{p}_k^2 | 0 \rangle \\
& + 4 \omega_0 \chi \chi' \langle 0 | \hat{\psi}_{k0}^2 | 0 \rangle \\
& - 2 \left( \xi \chi' + \chi \xi' \right) \\
& \times \langle 0 | \left( \hat{p}_{k0} \hat{\psi}_{k0} + \hat{\psi}_{k0} \hat{p}_{k0} \right) | 0 \rangle .
\end{split}
\end{equation}
Eqs. \eqref{eq:rho_k_mode}, \eqref{eq:psi_k_2_vac}, \eqref{eq:p_k_2_vac}, and \eqref{eq:psi_k_p_k_vac} explicitly expresses the terms $\langle 0 | \hat{ \rho }^{(\psi)} | 0 \rangle$ and $\langle 0 | \hat{ P }^{(\psi)} | 0 \rangle$ in the Friedmann equations (Eqs. \eqref{eq:fe1_conf} and \eqref{eq:fe2_conf}) in terms of the CHOs $\xi$ and $\chi$. Lastly, we prepare the initial $\hat{\psi}_k$-configuration in a Gaussian with a width $\sigma$, i.e., 
\begin{eqnarray}
\langle 0 | \hat{p}_{k0}^2 | 0 \rangle &=& \sigma / 2 \\
\langle 0 | \hat{\psi}_{k0}^2 | 0 \rangle &=& 1 / \left( 2 \sigma \right) \\
\langle 0 | \left( \hat{p}_{k0} \hat{\psi}_{k0} + \hat{\psi}_{k0} \hat{p}_{k0} \right) | 0 \rangle &=& 0 .
\end{eqnarray}
In this way, Eqs. \eqref{eq:psi_k_2_vac}, \eqref{eq:p_k_2_vac}, and \eqref{eq:psi_k_p_k_vac} further reduce to
\begin{eqnarray}
\langle 0 | \hat{\psi}_k^2 | 0 \rangle &=& \frac{\sigma}{\omega_0} \left( \xi^2 + \frac{ \omega_0^2 \chi^2 }{ \sigma^2 } \right) \\ 
\langle 0 | \hat{p}_k^2 | 0 \rangle &=& \frac{\sigma}{\omega_0} \left( \xi^{\prime 2} + \frac{ \omega_0^2 \chi^{\prime 2} }{\sigma^2} \right) \\
\langle 0 | \left( \hat{\psi}_k \hat{p}_k + \hat{p}_k \hat{\psi}_k \right) | 0 \rangle &=& \frac{ 2 \sigma }{\omega_0} \left( \xi \xi' + \frac{\omega_0^2}{\sigma^2} \chi \chi' \right) 
\end{eqnarray}
and the vacuum expectation values of $\hat{ \rho }_k^{(\psi)}$ and $\hat{ P }^{(\psi)}_k$ become
\begin{equation}
\begin{split}
\langle 0 | \hat{ \rho }_k^{(\psi)} | 0 \rangle 
= & \frac{1}{2 a^4} \bigg[
\frac{\sigma}{\omega_0} \left( \xi^{\prime 2} + \left( k^2 + \frac{a^{\prime 2}}{a^2} \right) \xi^2 - 2 \frac{a'}{a} \xi \xi' \right) \\
& + \frac{\omega_0}{\sigma} \left( \chi^{\prime 2} + \left( k^2 + \frac{a^{\prime 2}}{a^2} \right) \chi^2 - 2 \frac{a'}{a} \chi \chi' \right) \bigg] .
\end{split}
\end{equation}

Some final notes are in order to guarantee a well-posed initial value problem. First, by using Eq. \eqref{eq:omega_eta} and the Friedmann equations (Eqs. \eqref{eq:fe1_conf} and \eqref{eq:fe2_conf}), we can express $\omega_0$ as
\begin{equation}
\label{eq:omega_0_2}
\omega_0^2 = k^2 - \frac{a_0^2}{3} \left[ 2 \rho - \langle 0 | \hat{\rho}^{(\psi)} | 0 \rangle_0 \right]
\end{equation}
where $a_0$ and $\langle 0 | \hat{\rho}^{(\psi)} | 0 \rangle_0$ are the initial scale factor and the scalar field's energy density, respectively. The above equation also shows that the scalar field sources its own frequency -- a backreaction effect -- and that this effect is opposite to one where only the classical expansion drives the quantum field. It can also be shown that the initial value of the vacuum expectation values of the energy density and pressure of the modes are given by
\begin{equation}
\label{eq:rho_k_vac_0}
\langle 0 | \hat{ \rho }^{(\psi)}_k | 0 \rangle_0 = \langle 0 | \hat{ P }^{(\psi)}_k | 0 \rangle_0 = \frac{\sigma}{4 a_0^4} \left[ 1 + \frac{ k^2 + \left( a_0^{\prime}/a_0 \right)^2 }{\sigma^2} \right] .
\end{equation}

\subsubsection*{Summary}

The Einstein-(massless) Klein-Gordon system for a $k$-mode of the quantum scalar field that is initially Gaussian distributed with a width $\sigma$ is equivalent to the coupled system
\begin{equation}
\label{eq:friedmann_eq}
\begin{split}
3 \frac{a^{\prime 2}}{a^4} = \rho & +  \frac{1}{2 a^4} \bigg[  \frac{\sigma}{\omega_0} \left( \xi^{\prime 2} + \left( k^2 + \frac{a^{\prime 2}}{a^2} \right) \xi^2 - 2 \frac{a'}{a} \xi \xi' \right) \\ 
& \phantom{gg} + \frac{\omega_0}{\sigma} \left( \chi^{\prime 2} + \left( k^2 + \frac{a^{\prime 2}}{a^2} \right) \chi^2 - 2 \frac{a'}{a} \chi \chi' \right) \bigg] ,
\end{split}
\end{equation}
\begin{equation}
\label{eq:hubble_eq}
\begin{split}
2 \frac{a''}{a^3} - \frac{a^{\prime 2}}{a^4} = & -P - \frac{1}{2 a^4} \bigg[ 
\frac{\sigma}{\omega_0} \bigg( \xi^{\prime 2} - 2 \frac{a'}{a} \xi \xi' \\
& \phantom{ggggggggggggggggg} + \left( k^2 + \frac{a^{\prime 2}}{a^2} \right) \xi^2
\bigg) \\
& \phantom{gggg} + \frac{\omega_0}{\sigma} \bigg( \chi^{\prime 2} - 2 \frac{a'}{a} \chi \chi' \\
& \phantom{ggggggggggggg} + \left( k^2 + \frac{a^{\prime 2}}{a^2} \right) \chi^2 \bigg) \bigg] ,
\end{split}
\end{equation}
\begin{equation}
\label{eq:xi_cho_eom}
\xi'' + \omega^2 \left(\eta\right) \xi = 0 ,
\end{equation}
and
\begin{equation}
\label{eq:chi_cho_eom}
\chi'' + \omega^2 \left(\eta\right) \chi = 0
\end{equation}
where the driving frequency $\omega \left( \eta \right)$ is given by
\begin{equation}
\label{eq:omega_k_a}
\omega^2 \left( \eta \right) = k^2 - \frac{a''}{a} 
\end{equation}
and the initial conditions are
\begin{eqnarray}
\label{eq:a0} a \left( \eta_0 \right) &=& a_0 \\
\label{eq:ap0} a' \left( \eta_0 \right) &=& a_0' \\
\label{eq:xi0} \xi \left( \eta_0 \right) &=& 0 \\ 
\label{eq:xip0} \xi' \left( \eta_0 \right) &=& \sqrt{\omega_0 / 2} \\
\label{eq:chi0} \chi \left( \eta_0 \right) &=& - 1/ \sqrt{2\omega_0} \\
\label{eq:chip0} \chi' \left( \eta_0 \right) &=& 0 .
\end{eqnarray}

\subsection{Numerical investigation}
\label{subsec:numerical_investigation}

The numerical execution is described as follows. We combine the Friedmann constraint (Eq. \eqref{eq:friedmann_eq}) with the Hubble equation (Eq. \eqref{eq:hubble_eq}) to obtain
\begin{equation}
\label{eq:fe12_cqc}
\begin{split}
\frac{a''}{a} = \frac{2}{3} a^2 \rho &- \frac{1}{6 a^2} \bigg[
 \frac{\sigma}{\omega_0} \left( \xi^{\prime 2} + \left( k^2 + \frac{a^{\prime 2}}{a^2} \right) \xi^2 - 2 \frac{a'}{a} \xi \xi' \right) \\
&+ \frac{\omega_0}{\sigma}  \left( \chi^{\prime 2} + \left( k^2 + \frac{a^{\prime 2}}{a^2} \right) \chi^2 - 2 \frac{a'}{a} \chi \chi' \right) \bigg] .
\end{split}
\end{equation}
Also, we fix $\sigma = \omega_0$ throughout, thus, locking the initial state of the QHO to its instantaneous ground state. Previous implementations of the CQC have always invoked this. Remarkably, the main conclusion of this paper -- the existence of a short wavelength threshold for inflationary stability -- is valid even for $\sigma \neq \omega_0$. Eqs. \eqref{eq:xi_cho_eom}, \eqref{eq:chi_cho_eom}, and \eqref{eq:fe12_cqc} are then solved for $\left( a, \xi, \chi \right)$ provided the initial conditions given by Eqs. \eqref{eq:a0}, \eqref{eq:ap0}, \eqref{eq:xi0}, \eqref{eq:xip0}, \eqref{eq:chi0}, and \eqref{eq:chip0}. To benchmark the results with the classical limit, i.e., $\langle 0 | \hat{\rho}^{(\psi)} | 0 \rangle, \langle 0 | \hat{P}^{(\psi)} | 0 \rangle \rightarrow 0$, we choose the initial conditions of the spacetime to match the de Sitter solution,
\begin{equation}
\label{eq:a_ds_classical}
a \left( \eta \right) = \frac{a_0}{1 - a_0 h \eta} ,
\end{equation}
with a Hubble parameter $H = a'/a^2 = h$. The independent parameters of the system are treated as $k$, $a_0$, and $h$. The initial state of the CHOs (Eqs. \eqref{eq:xi0}, \eqref{eq:xip0}, \eqref{eq:chi0}, and \eqref{eq:chip0}) can be computed using these parameters, i.e., calculate $\omega_0$ using Eqs. \eqref{eq:omega_0_2} and \eqref{eq:rho_k_vac_0}.

\begin{figure}[h!]
\center
\includegraphics[width = 0.4 \textwidth]{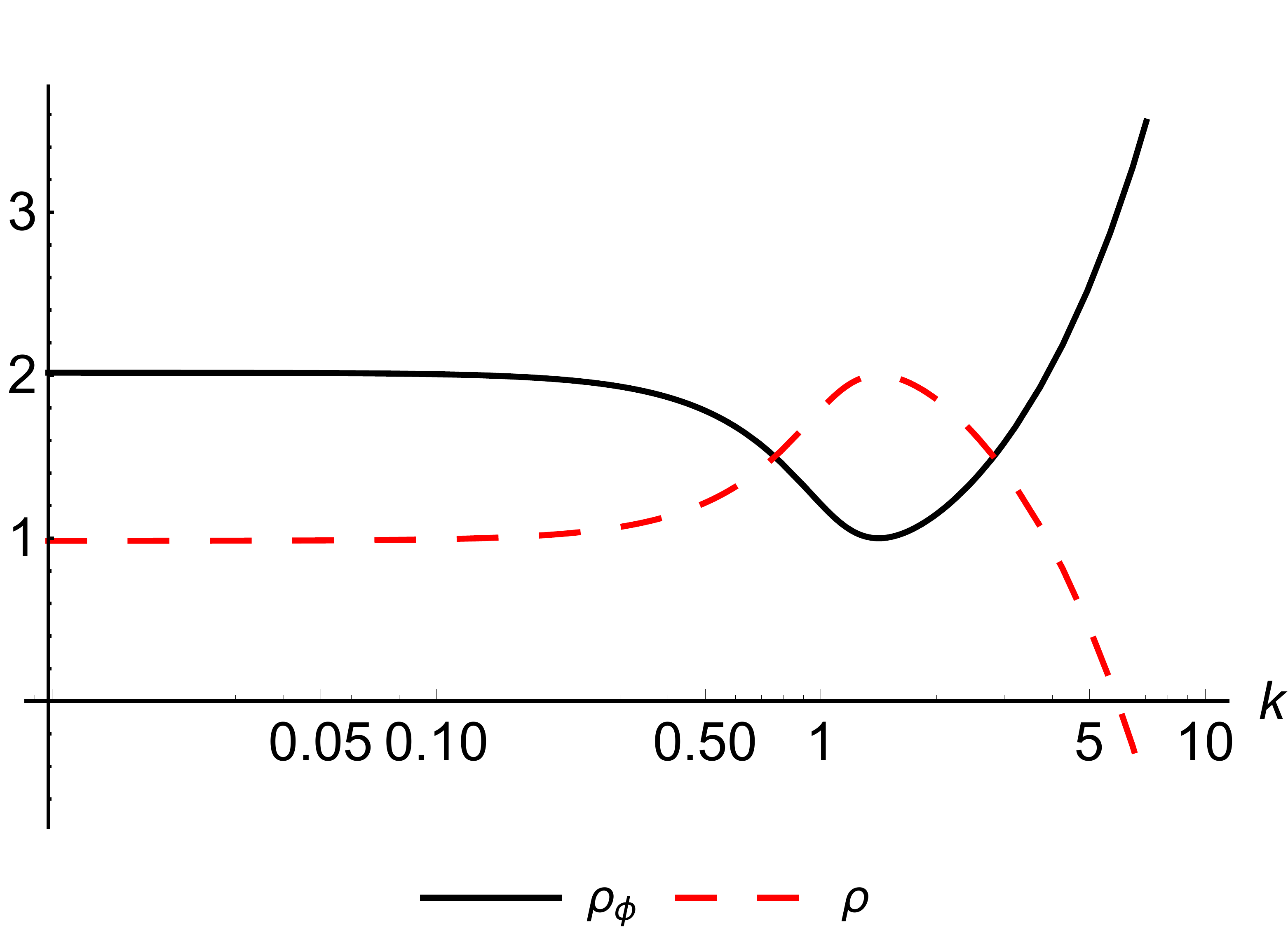}
\caption{Initial energy densities of the scalar field, $\rho_\phi \left( k \right) = \langle 0 | \hat{\rho}^{\left(\psi\right)} | 0 \rangle$, and vacuum energy, $\rho \left( k \right)$, as a function of the mode's wavenumber $k$ for $a_0 = 1$ and $h = 1$. The sum of the densities is fixed $\left( \rho_\phi + \rho = 3 \right)$ because of the Friedmann constraint. Numerical inspection shows that $\rho_\phi \left( k \ll 1 \right) = 2.02$, $\rho_\phi \left( k \approx 1.41 \right) = 1.00$, and $\rho_\phi \left( k \approx 3.90 \right) = 2.02$ where $k \approx 1.41$ marks the minimum of $\rho_\phi \left( k \right)$ and $k \approx 3.90$ signify the wavenumber above which $\rho_\phi \left( k \right) > 2.02$.}
\label{fig:rho_0}
\end{figure}

It is important to also note that by choosing $k$, $a_0$, and $h$ as independent parameters we have foregone control over the vacuum energy density $\rho$, i.e., $\rho = \rho \left( k \right)$ for fixed $a_0$ and $h$. This dependence is transparently revealed in figure \ref{fig:rho_0}.
A numerical investigation of this shows that the scalar field's energy density, $\rho_\phi = \langle 0 | \hat{\rho}^{\left(\psi\right)} | 0 \rangle$, has a minimum value of $\rho_\phi \approx 1.00$ at $k \approx 1.41$ and that $\rho_\phi < 2.02$ for $0 \leq k < 3.90$.
It is also worth pointing out that the vacuum energy, $\rho$, assumes negative values for $k \gtrapprox 5.92$. The destabilization of the inflationary era might therefore be naively associated with this. However, it will be shown for the nearly critical wavelengths ($k \sim 5.78$) that this does not appear to be a straightforward conclusion. The succeeding discussion is hereby divided into long wavelength-modes (Section \ref{subsubsec:long}), nearly critical wavelength-modes (Section \ref{subsubsec:crit}), and short wavelength-modes (Section \ref{subsubsec:short}).

\subsubsection{Long wavelength-modes}
\label{subsubsec:long}

We present results of the integration for long wavelength-modes. 
\begin{figure*}[h!]
\center
	\subfigure[ scale factor ]{
		\includegraphics[width = 0.4 \textwidth]{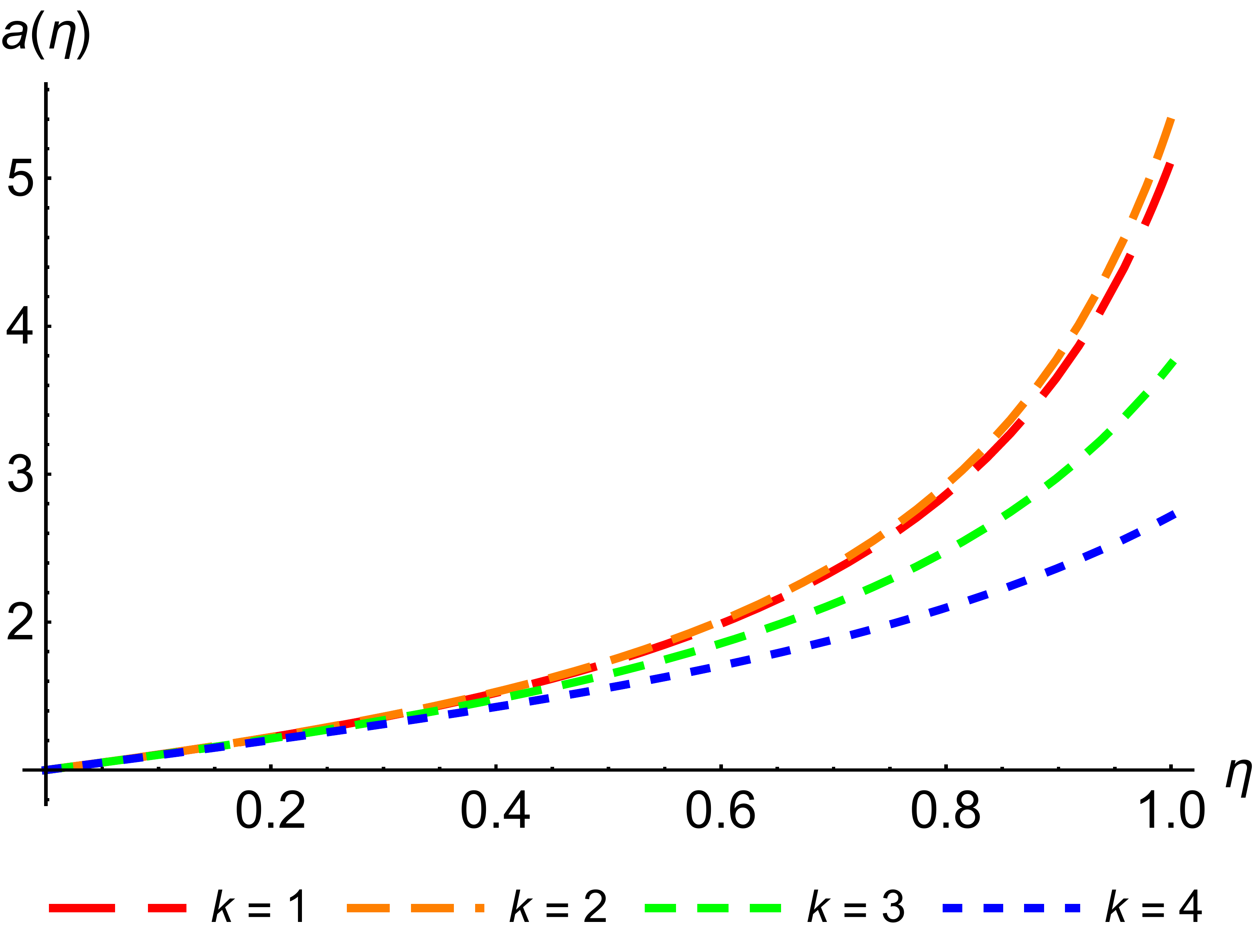}
		}
	\subfigure[ Hubble parameter ]{
		\includegraphics[width = 0.4 \textwidth]{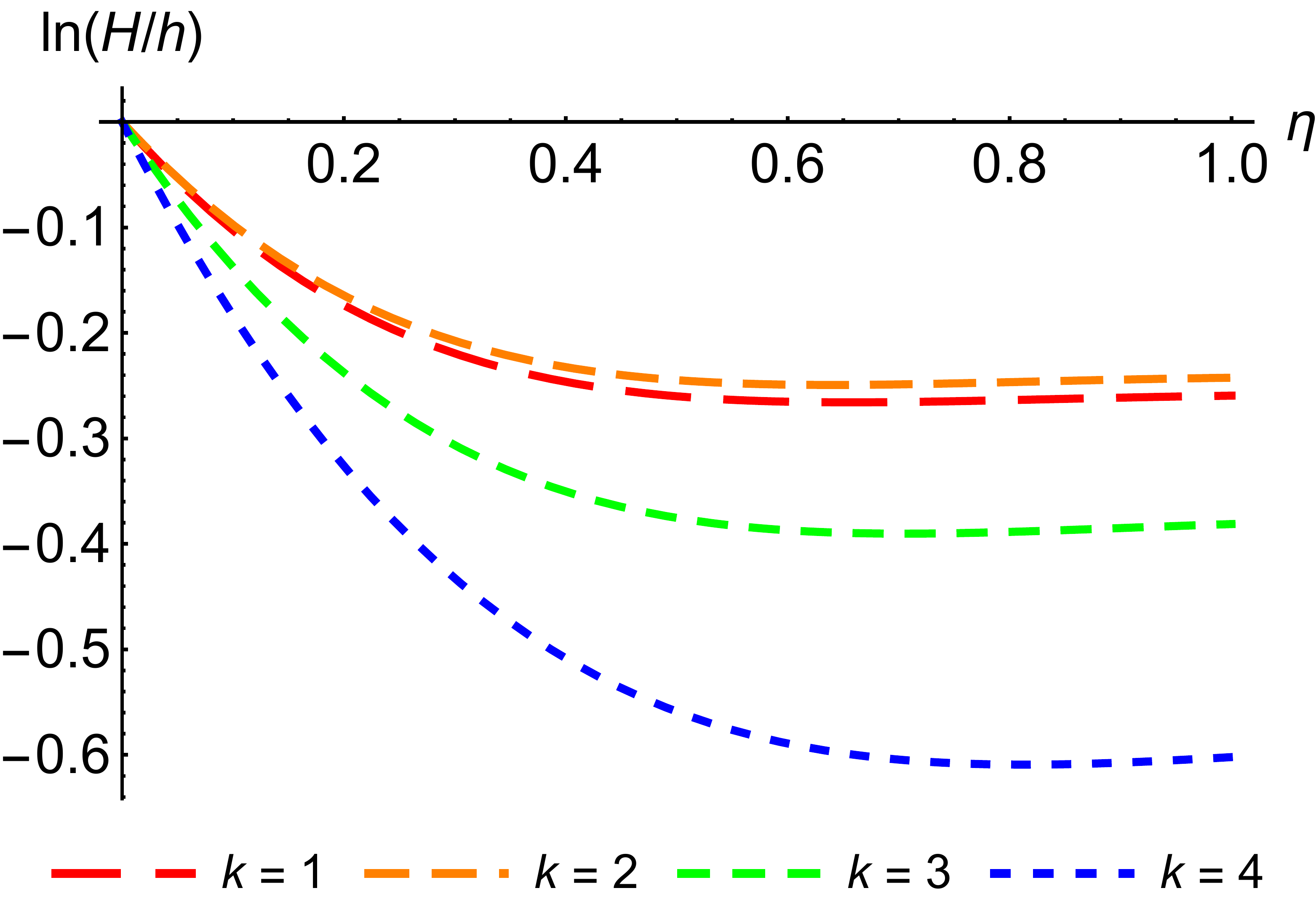}
		}
	\subfigure[ QHO Hamiltonian ]{
		\includegraphics[width = 0.4 \textwidth]{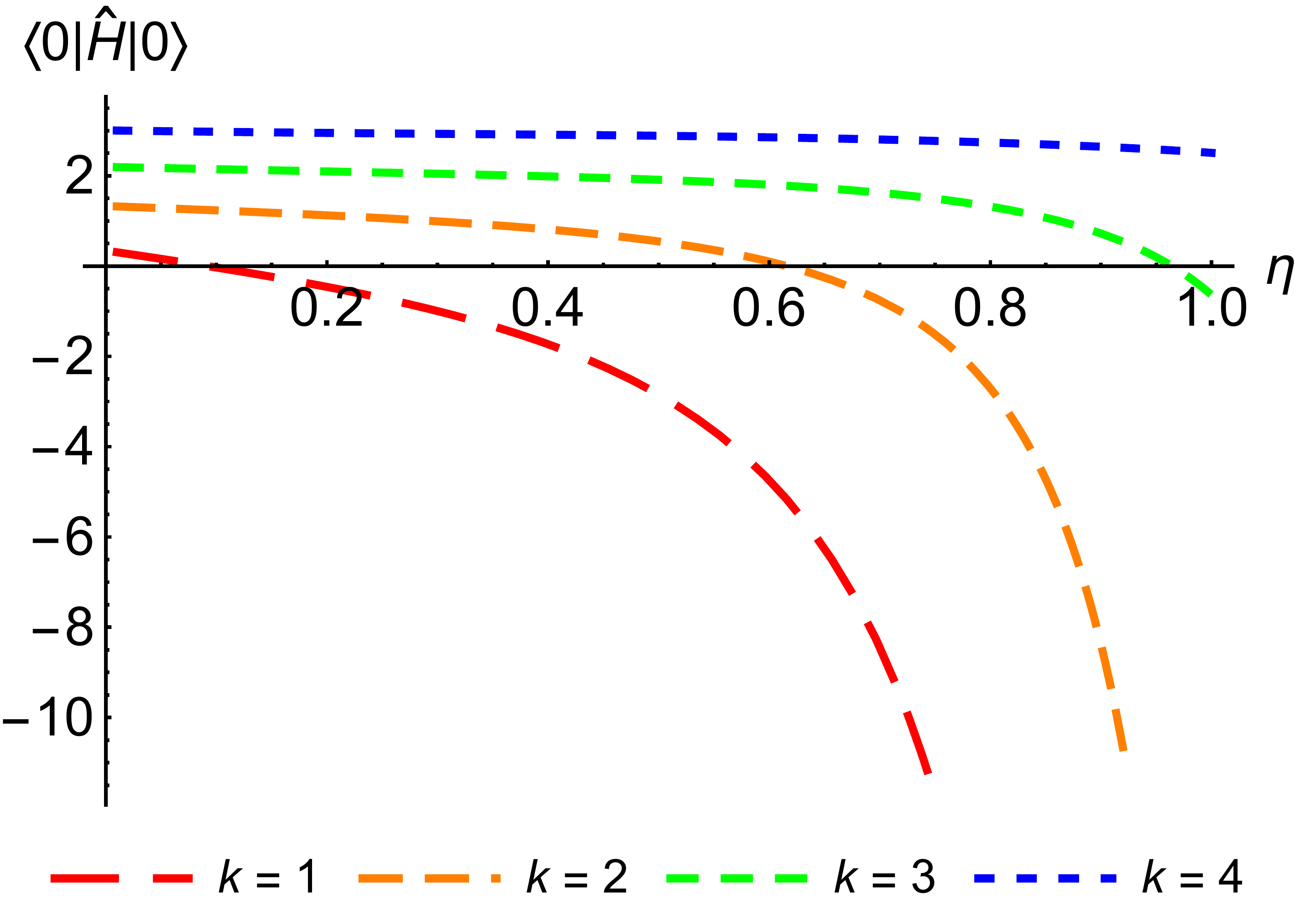}
		}
	\subfigure[ frequency squared ]{
		\includegraphics[width = 0.4 \textwidth]{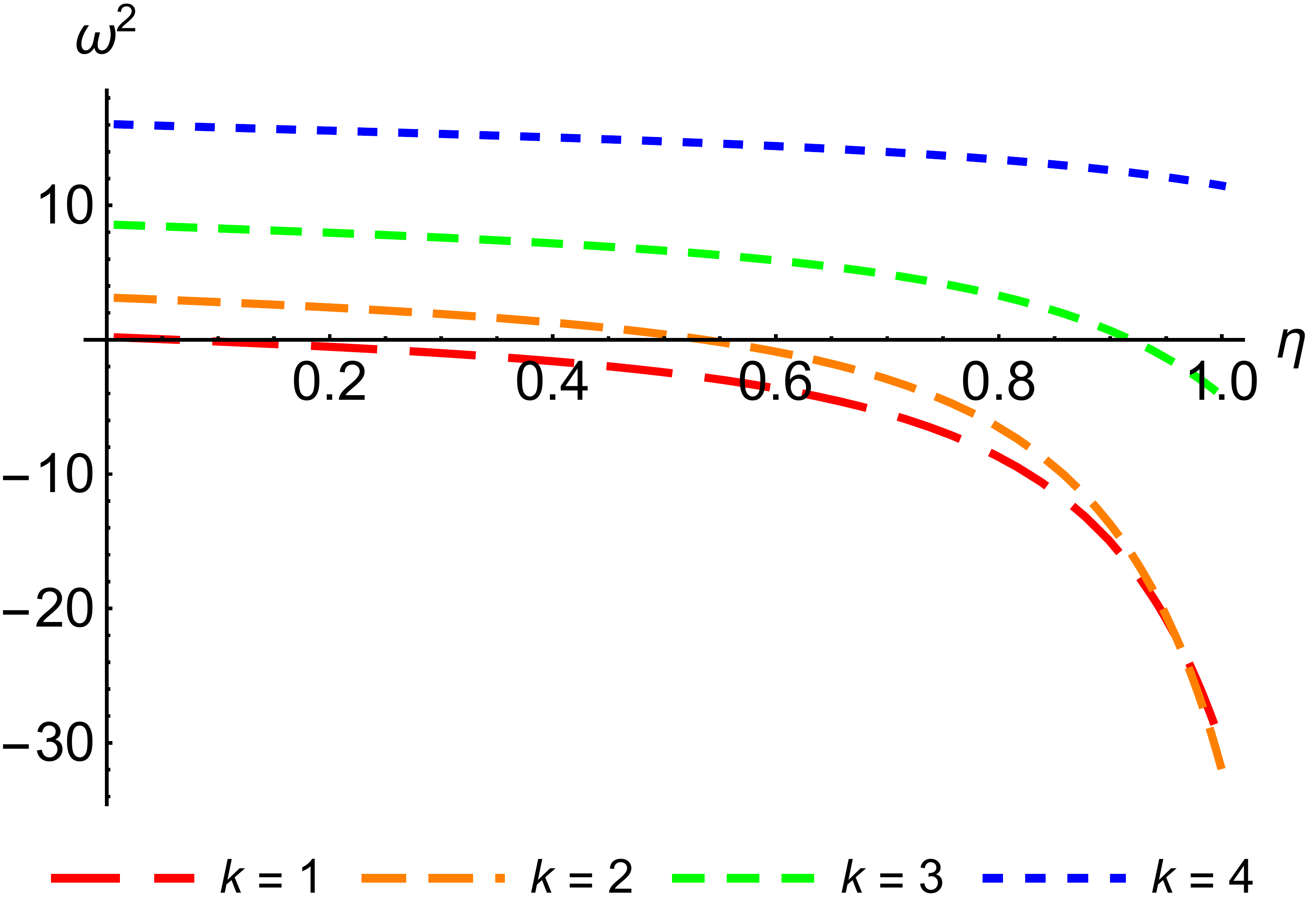}
		}
	\subfigure[ particle number ]{
		\includegraphics[width = 0.4 \textwidth]{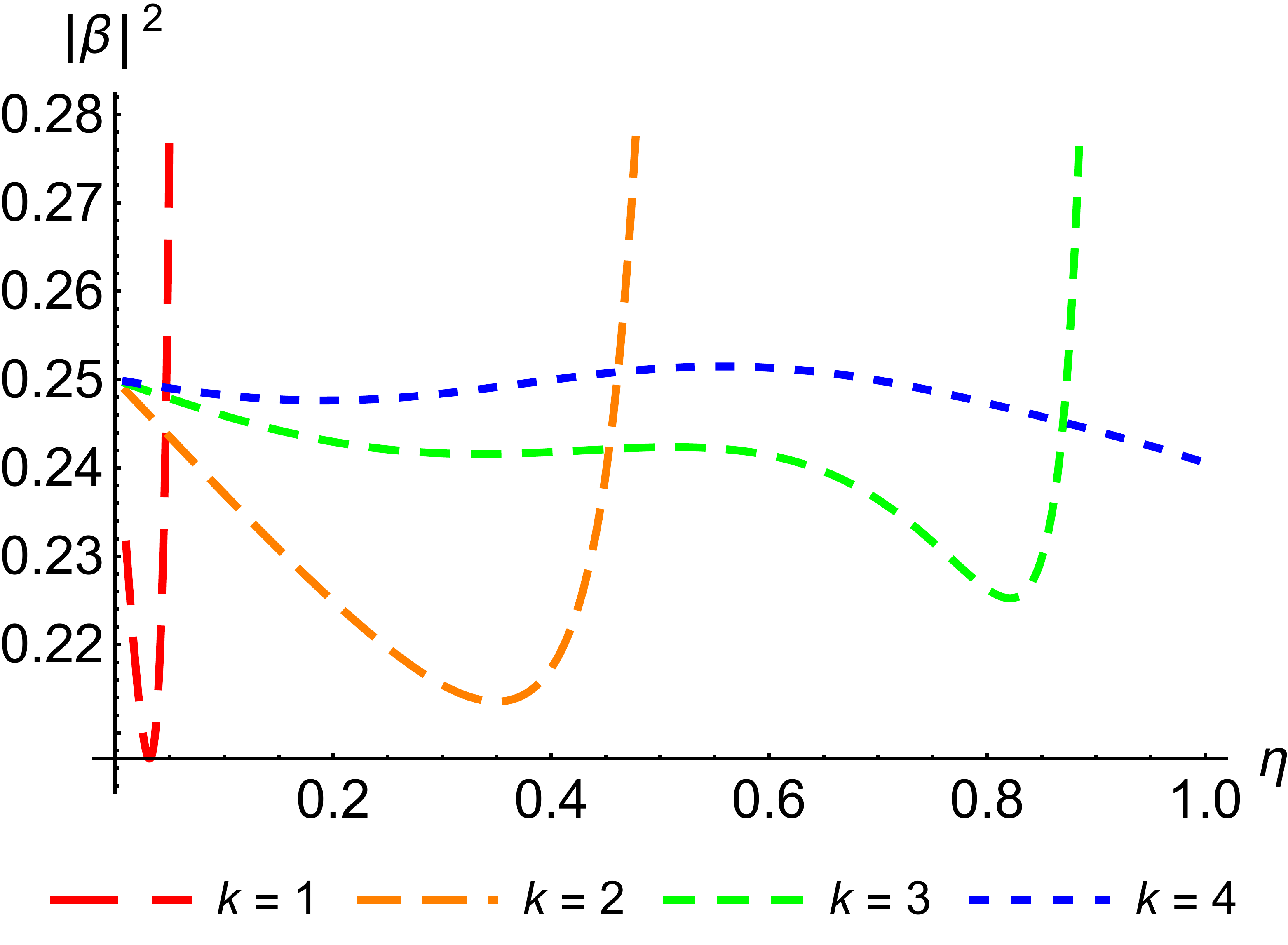}
		}
	\subfigure[ scalar field's energy density ]{
		\includegraphics[width = 0.4 \textwidth]{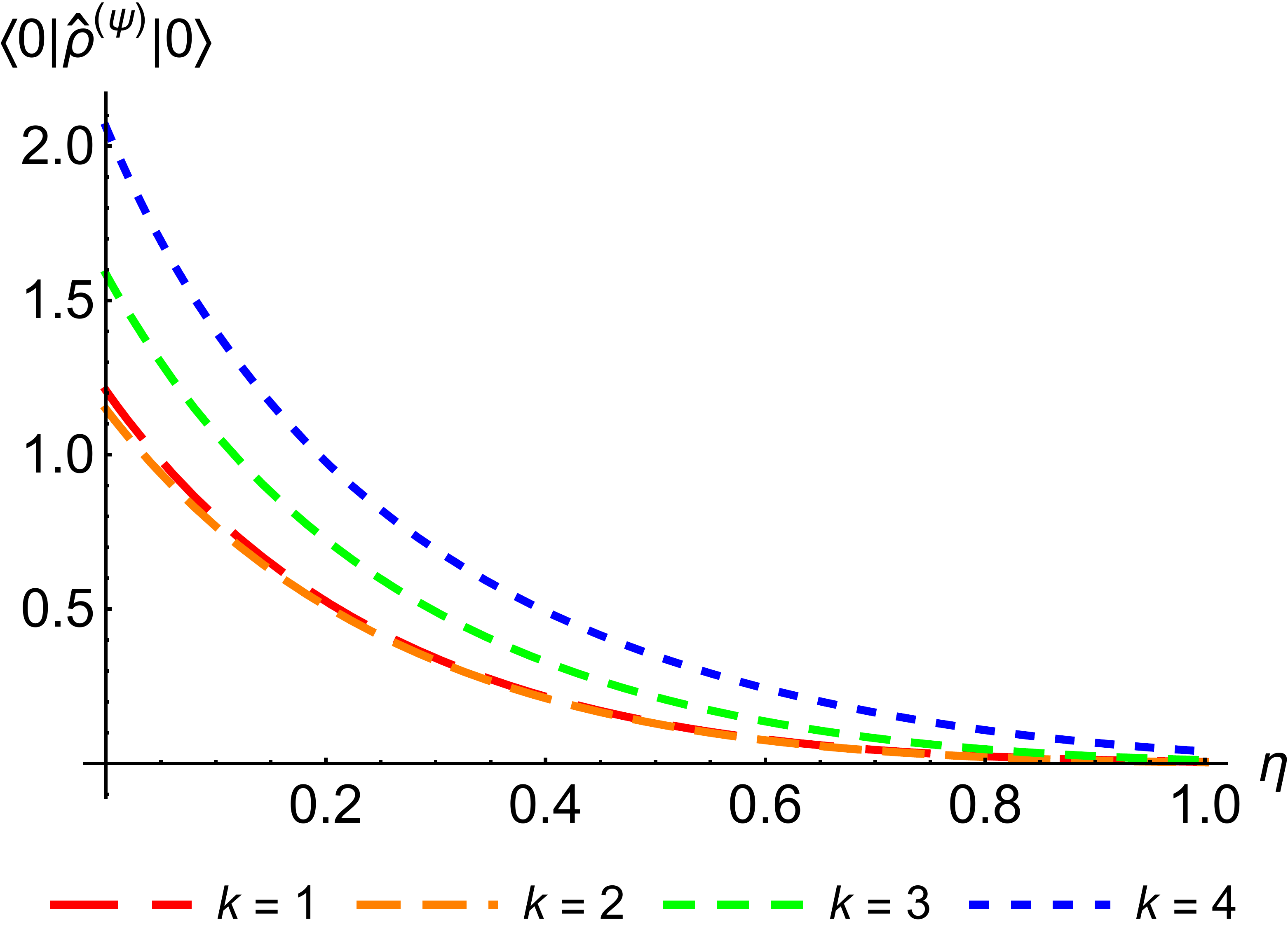}
		}
\caption{Plots of (a) scale factor, $a$, (b) Hubble parameter, $H = a'/a^2$, (c) QHO Hamiltonian, $\langle 0 | \hat{H} | 0 \rangle$, (d) frequency squared, $\omega^2$, (e) particle number, $| \beta |^2$, and (f) scalar field's energy density, $\langle 0 | \hat{\rho}^{\left(\psi\right)} | 0 \rangle$, obtained using the CQC for $k = 1, 2, 3, 4$ with $a_0 = 1$ and $h = 1$.}
\label{fig:long}
\end{figure*}
Figures \ref{fig:long}(a) and (b) show that the background spacetime eventually ends up with an exponential expansion. This is more convincingly observed in figure \ref{fig:long}(b) where the logarithm of the Hubble parameter $( H = a'/a^2 )$ undoubtedly becomes constant, marking the onset of (a \textit{renormalized}) inflationary era, during the late stages of the evolution. The backreaction can indeed be confirmed to renormalize the Hubble parameter to a smaller value $H_\text{eff} < h$, where $H_\text{eff} = H \left( \eta \rightarrow \eta_f \right)$ with $\eta_f$ corresponding to $t \rightarrow \infty$, for all $k < 5.78$. This defines the long wavelength regime. In addition, figures \ref{fig:long}(a) and (b) explicitly show that $k = 1$ backreacts to the spacetime more strongly compared to $k = 2$, i.e., $H_{\text{eff} , k = 1} < H_{ \text{eff} , k = 2} < h$. This can be explained by the minimum value of $\rho_\phi$ for $k \approx 1.41$ (figure \ref{fig:rho_0}) and in particular because $\rho_\phi \left( k = 1 \right) > \rho_\phi \left( k = 2 \right)$, noting that $\rho_\phi \left( k \right)$ can be taken to quantify the strength of backreaction of a mode $k$. The results of the integration, $\left( a , \xi, \chi \right)$, could of course be used to compute any desired physical quantity. Shown in figures \ref{fig:long}(c), (d), (e), and (f) are therefore the corresponding QHO Hamiltonian $\left( \langle 0 | \hat{H} | 0 \rangle \right)$, frequency squared $\left(\omega^2\right)$, particle number $\left( | \beta |^2 \right)$, and scalar field's energy density $\left( \langle 0 | \hat{\rho}^{\left(\psi\right)} | 0 \rangle \right)$, respectively. It can be seen that the
QHO's Hamiltonian (figure \ref{fig:long}(c)) as well as the square of the frequency (figure \ref{fig:long}(d)) follow a decreasing trend in their evolution, even crossing to negative values for the longer wavelengths. The trail to negative Hamiltonian shows that the long wavelength modes are themselves unstable and so they must interact with other fields later through some unaccounted mechanism. Nonetheless, particle creation continues to occur as shown in figure \ref{fig:long}(e). Mode production, marked by the eventual rise in the particle number, $| \beta |^2$, occurs earlier for the longer wavelength cases, as they should be easier to pull out of the vacuum. This is clearly the case in figure \ref{fig:long}(e) where $|\beta|^2$ begins to exponentially rise earlier for $k_a$ compared to $k_b > k_a$, e.g., for $k = 1$ the onset of mode production occurs near $\eta = 0.04$ while for $k = 2$ this occurs near $\eta = 0.36$. The scalar field's energy density, $\langle 0 | \hat{\rho}^{\left(\psi\right)} | 0 \rangle$, declines despite this because of the ongoing expansion and anticipates the upcoming inflationary era (figure \ref{fig:long}(f)). 

\subsubsection{Near threshold wavelength-modes}
\label{subsubsec:crit}

We present the results of the integration of the field equations for nearly critical wavelength-modes, i.e., modes characterized by wavelengths very close to the threshold. 
\begin{figure*}[h!]
\center
	\subfigure[ scale factor ]{
		\includegraphics[width = 0.4 \textwidth]{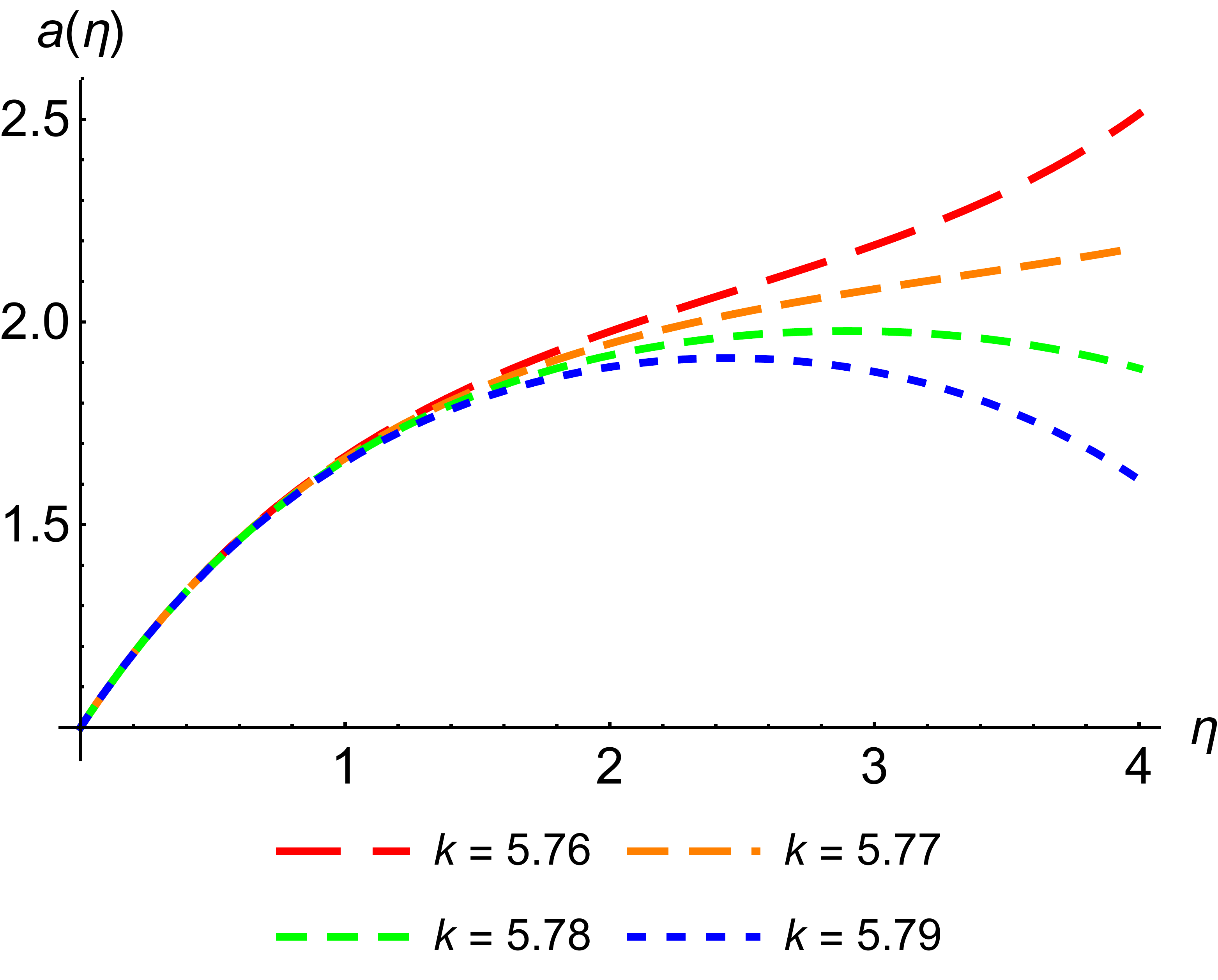}
		}
	\subfigure[ Hubble parameter ]{
		\includegraphics[width = 0.4 \textwidth]{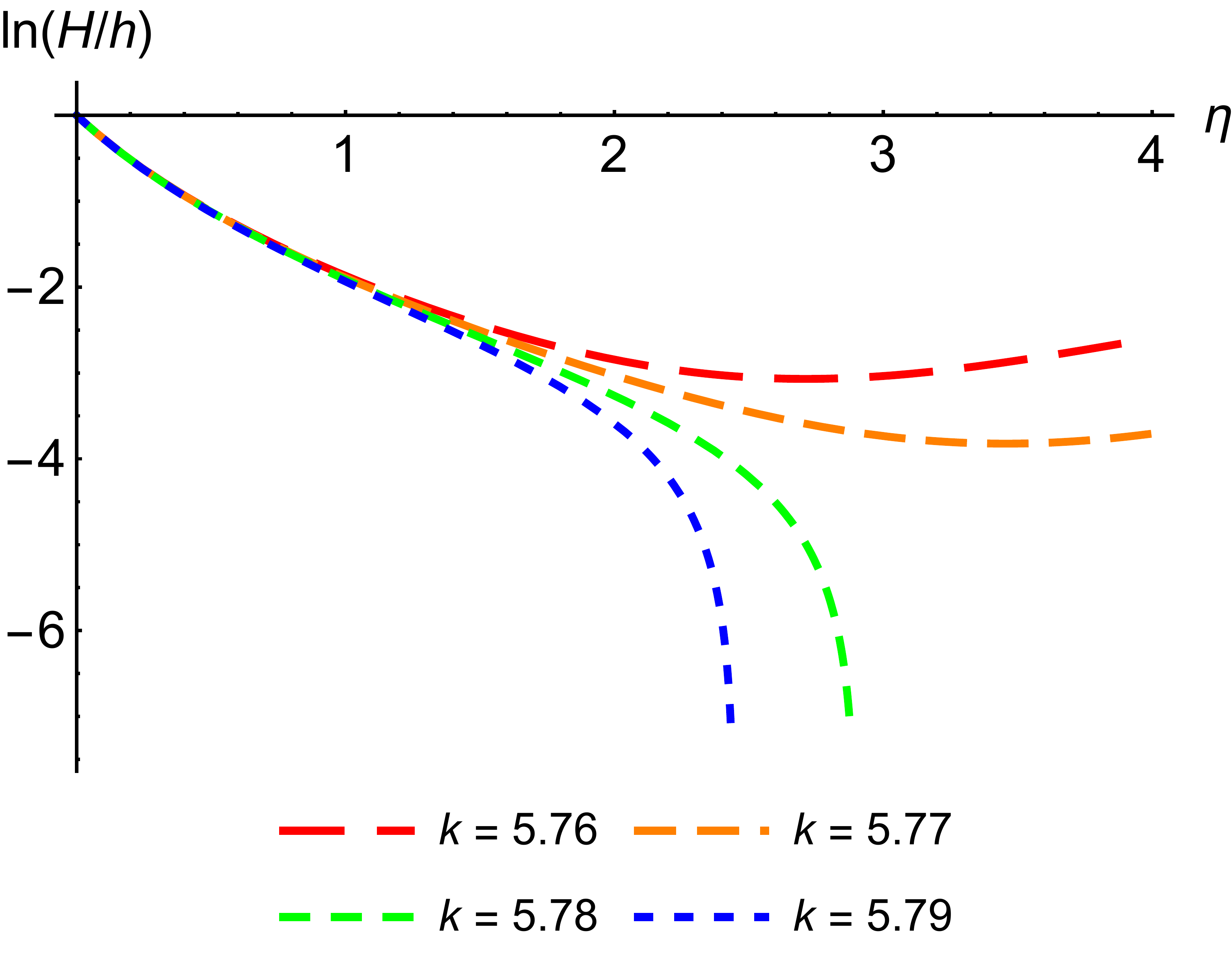}
		}
	\subfigure[ QHO Hamiltonian ]{
		\includegraphics[width = 0.4 \textwidth]{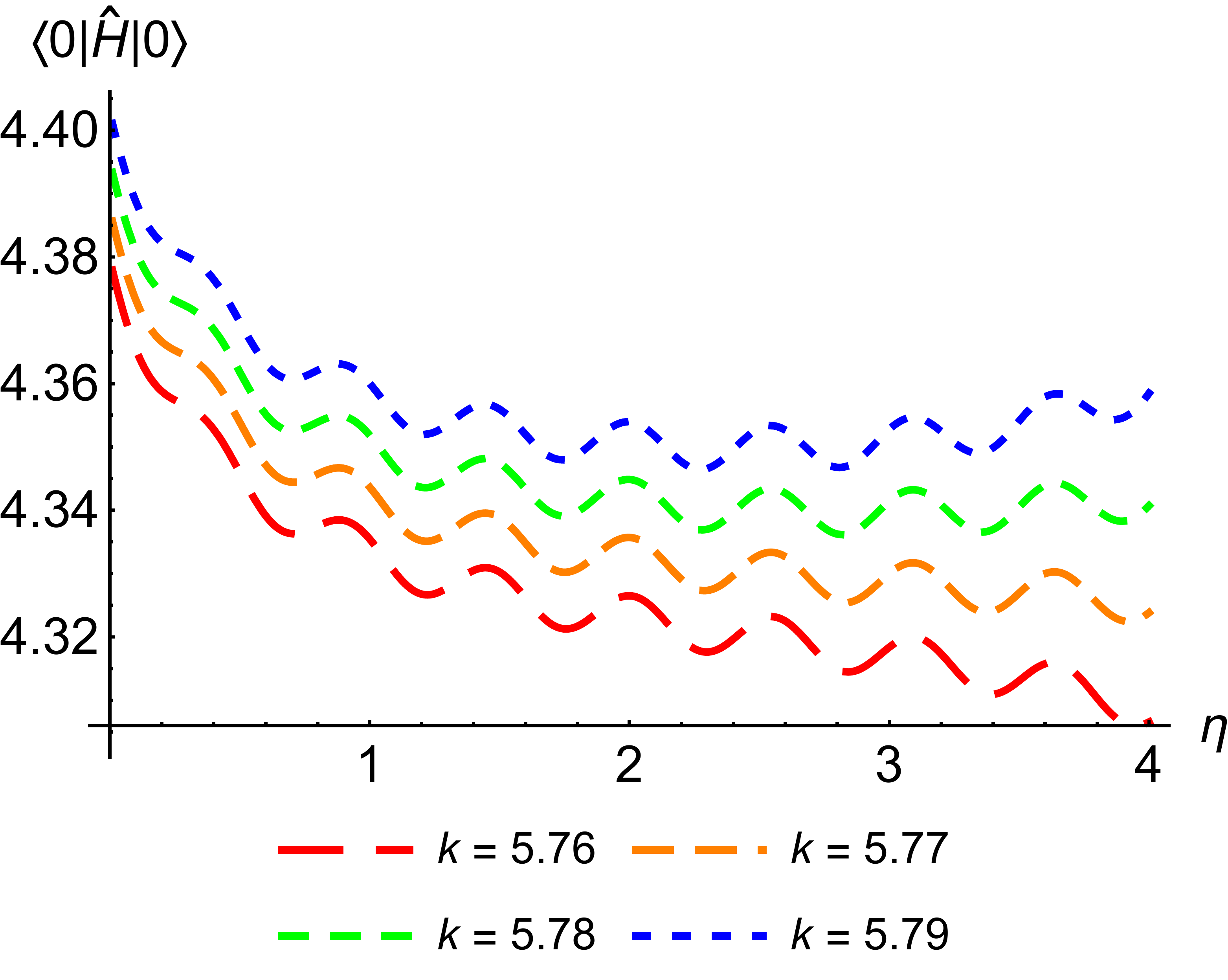}
		}
	\subfigure[ frequency squared ]{
		\includegraphics[width = 0.4 \textwidth]{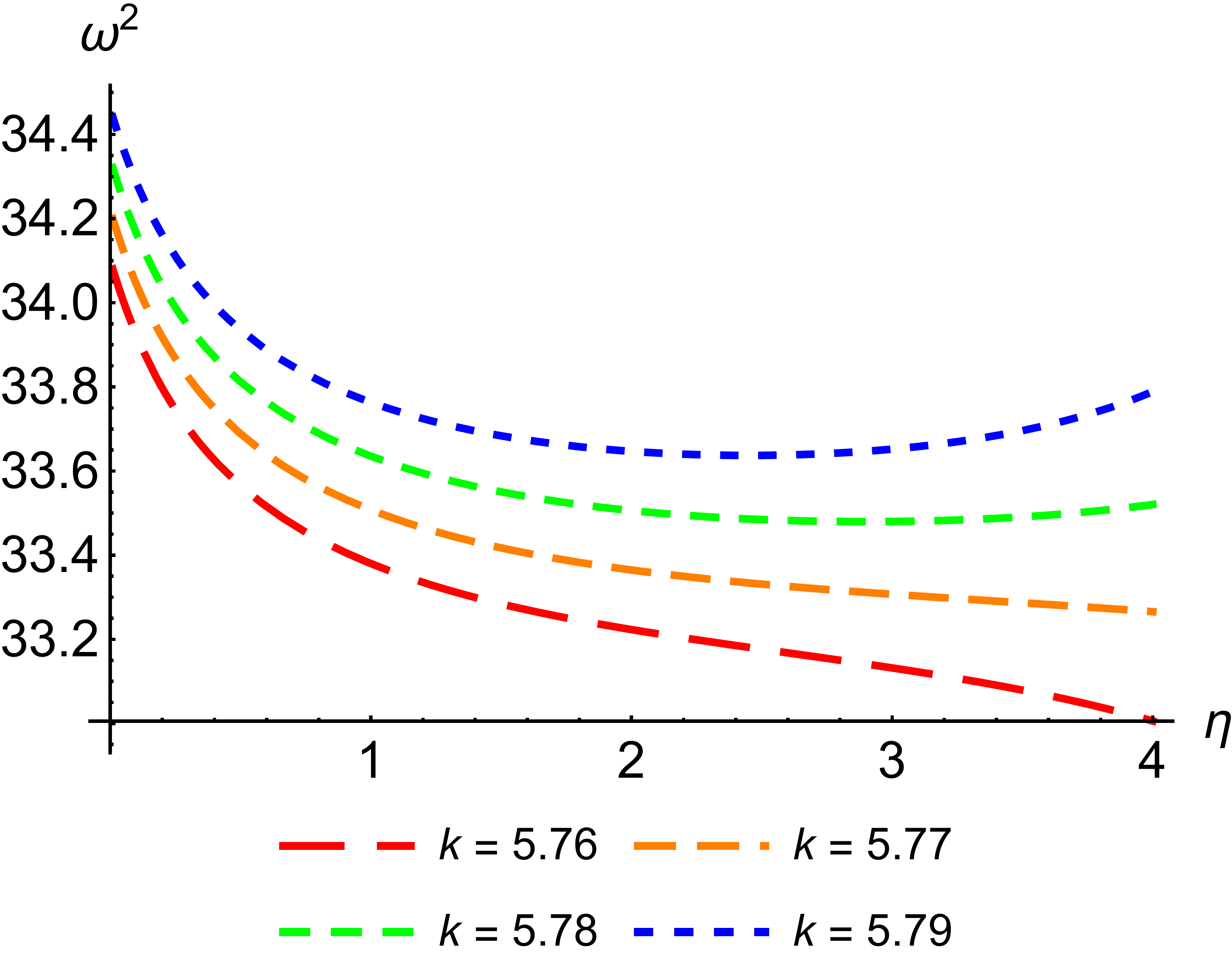}
		}
	\subfigure[ particle number ]{
		\includegraphics[width = 0.4 \textwidth]{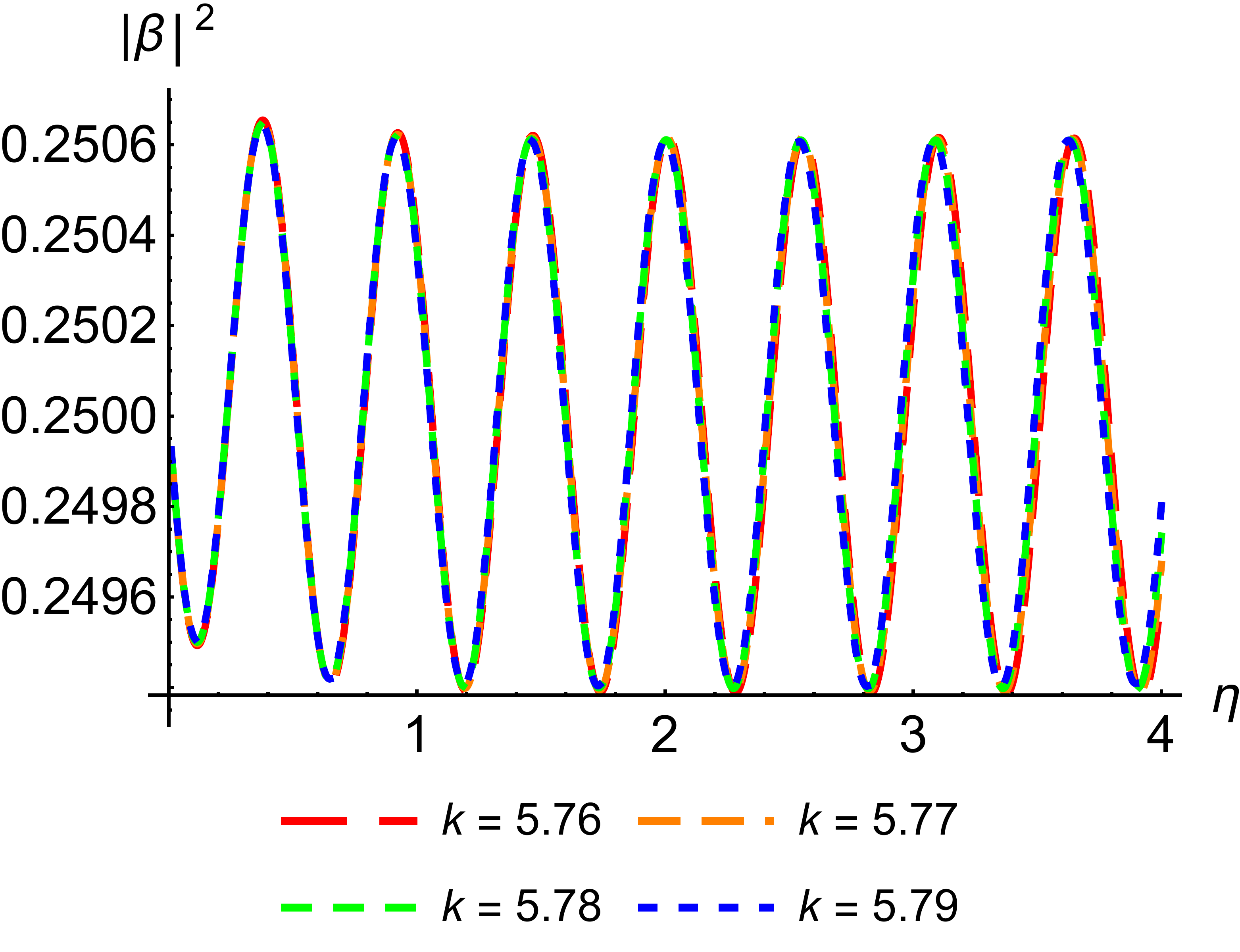}
		}
	\subfigure[ scalar field's energy density ]{
		\includegraphics[width = 0.4 \textwidth]{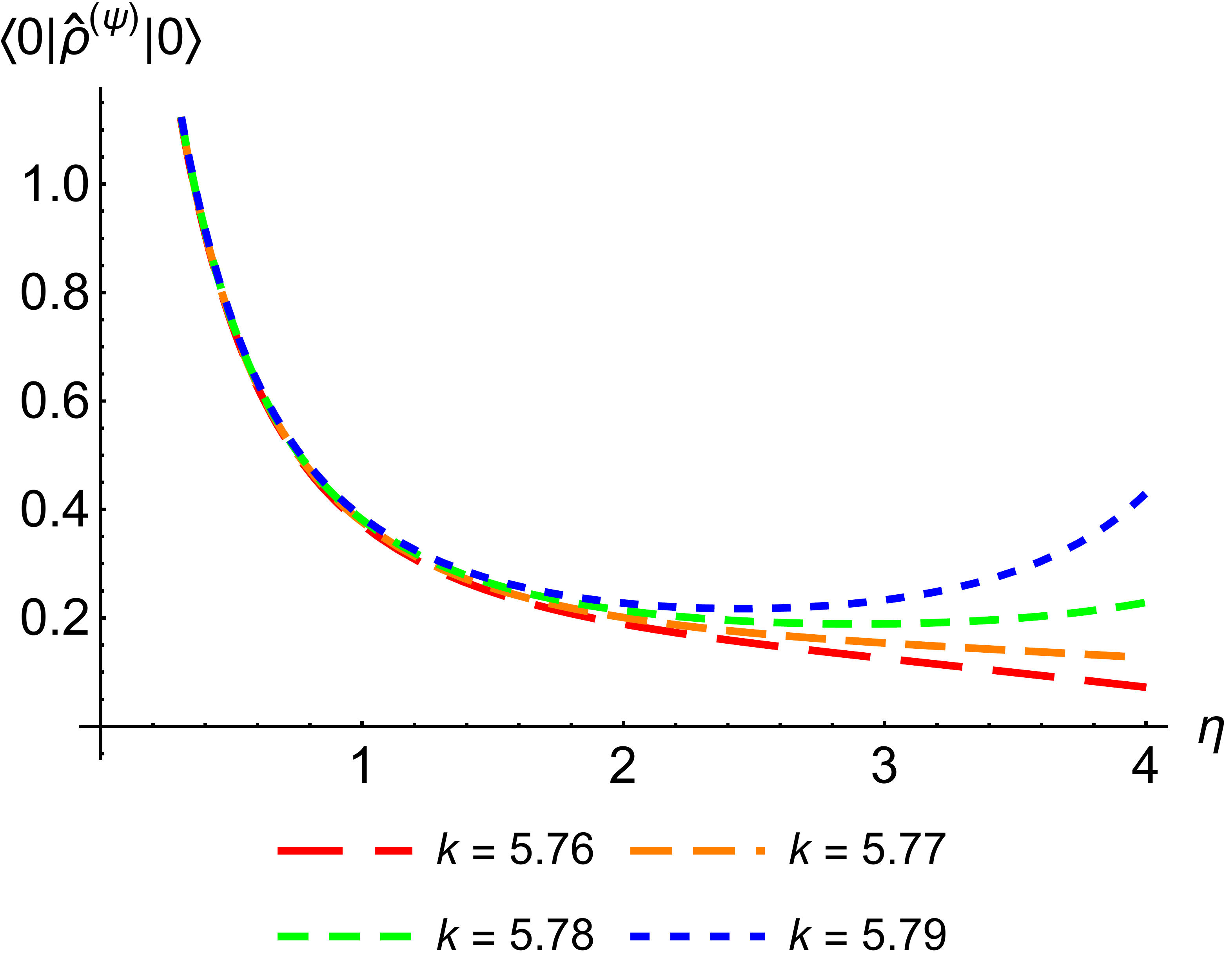}
		}
\caption{Plots of (a) scale factor, $a$, (b) Hubble parameter, $H = a'/a^2$, (c) QHO Hamiltonian, $\langle 0 | \hat{H} | 0 \rangle$, (d) frequency squared, $\omega^2$, (e) particle number, $| \beta |^2$, and (f) scalar field's energy density, $\langle 0 | \hat{\rho}^{\left(\psi\right)} | 0 \rangle$, obtained using the CQC for $k = 5.76, 5.77, 5.78, 5.79$ with $a_0 = 1$ and $h = 1$.}
\label{fig:crit}
\end{figure*}
Figures \ref{fig:crit}(a) and (b) show the drastic shift in the behavior of the background, from inflationary to collapsing, upon crossing a certaint cutoff wavenumber, $k_\text{max} \approx 5. 77$. This defines the threshold wavelength $\lambda_{\text{min}} = 2\pi / k_\text{max}$ which conveniently differentiates the long wavelength-modes $\left(\lambda > \lambda_{\text{min}}\right)$ which will only tend to renormalize the Hubble parameter from short wavelength-modes $\left(\lambda < \lambda_{\text{min}}\right)$ that will cause the spacetime to collapse. It should be stressed out that $\rho > 0$ for all of the cases considered in figure \ref{fig:crit} and so the eventual collapse for the short wavelength-cases $\left(\lambda < \lambda_{\text{min}}\right)$ cannot be straightforwardly attributed to a negative value of vacuum energy. In particular, it can be shown by numerical inspection of figure \ref{fig:rho_0} that the nondynamical vacuum energy, $\rho$, will become negative only for $k \gtrapprox 5.92$. The near threshold wavelength-modes can also be seen to support a positive, slightly decreasing, Hamiltonian (figure \ref{fig:crit}(c)) and squared frequencies (figure \ref{fig:crit}(d)) throughout their evolution. They can therefore be considered to be more stable compared to the long wavelength-modes which cascade to negative energies. Particle creation, however, cannot anymore be observed to continue because of the stronger backreaction (figure \ref{fig:crit}(e)); instead, the particle number, $|\beta|^2$, just oscillates up and down for the near threshold-cases. The scalar field's energy density, $\langle 0 | \hat{\rho}^{\left(\psi\right)} | 0 \rangle$, on the other hand, increases during the late stages of the evolution for the nearly critical wavelength-modes with $\lambda < \lambda_{\text{min}}$ as  shown by careful inspection of figure \ref{fig:crit}(f) for $k = 5.78$ and $k = 5.79$. This is of course understood to be an effect of the collapsing era on the scalar field's energy density.

In a short while, it will be shown that the existence of the threshold wavelength can be crudely predicted analytically by assuming that the frequency $\omega$ is constant throughout (Section \ref{subsubsec:analytical_solution}). This rough approximation (Eq. \eqref{eq:a_short_wavelength}) holds well for very short wavelength-modes and supports the existence of a threshold.

\subsubsection{Short wavelength-modes}
\label{subsubsec:short}

We present the results of the integration for the short wavelength-modes.
\begin{figure*}[h!]
\center
	\subfigure[ scale factor ]{
		\includegraphics[width = 0.4 \textwidth]{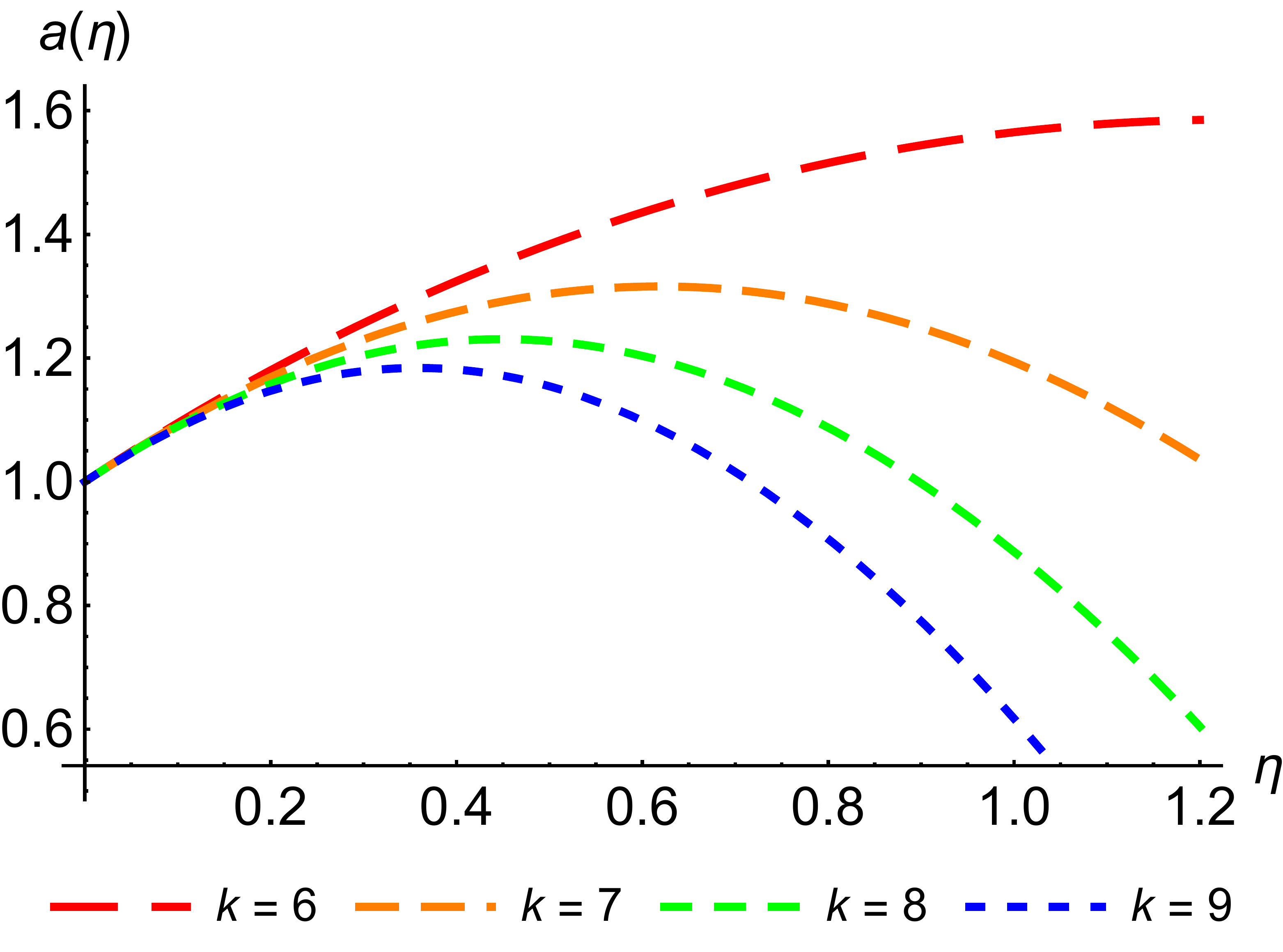}
		}
	\subfigure[ Hubble parameter ]{
		\includegraphics[width = 0.4 \textwidth]{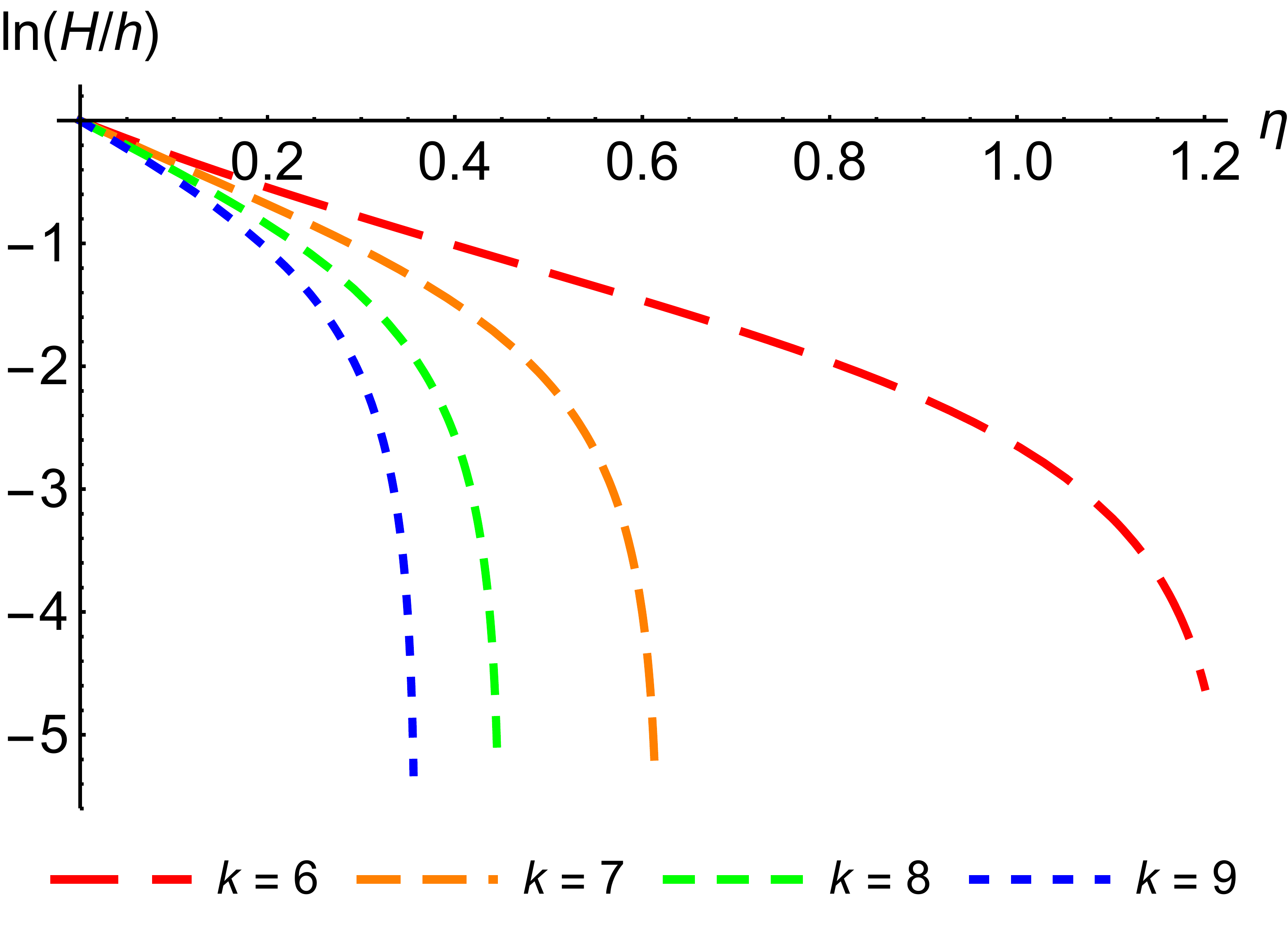}
		}
	\subfigure[ QHO Hamiltonian ]{
		\includegraphics[width = 0.4 \textwidth]{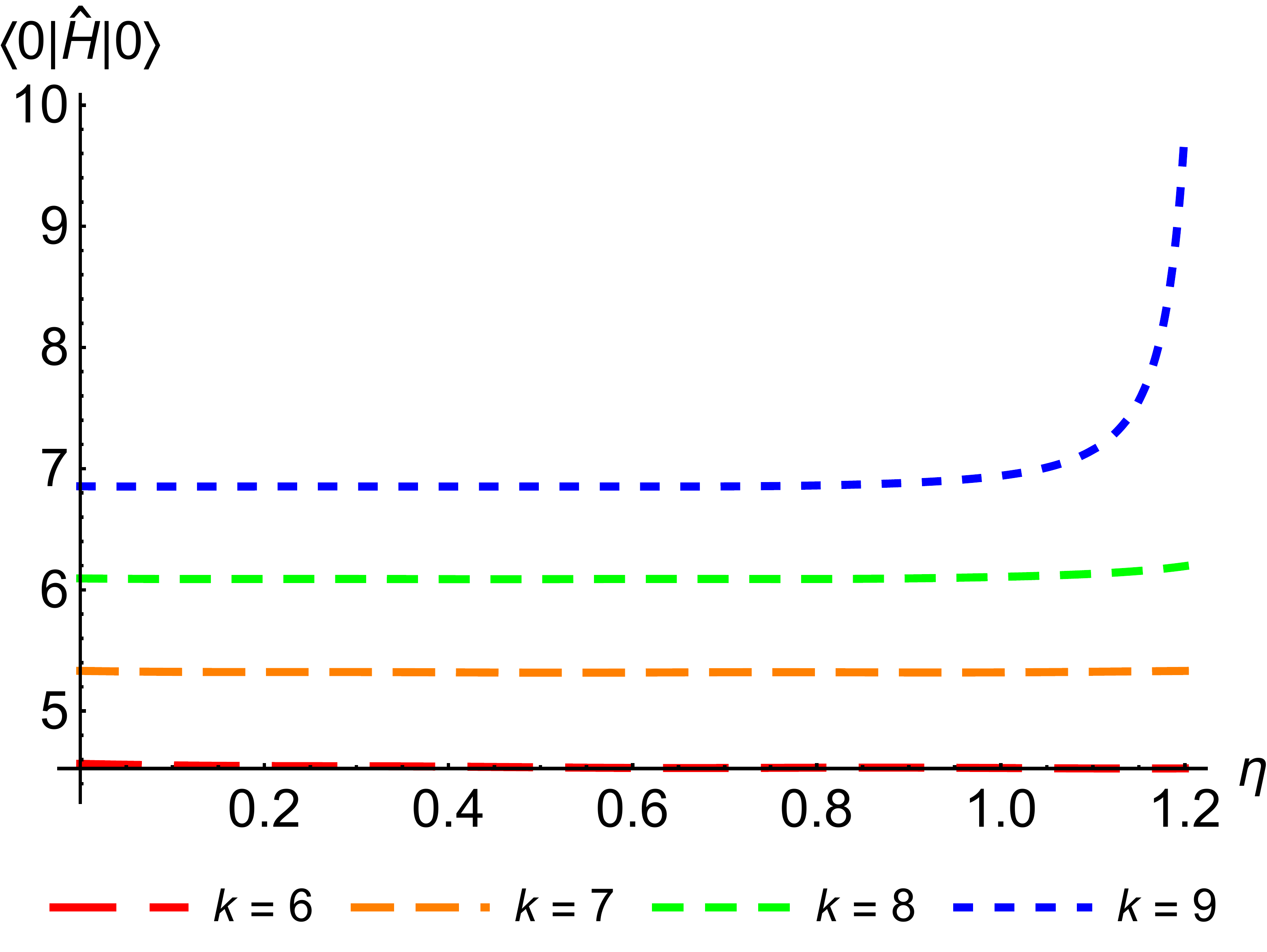}
		}
	\subfigure[ frequency squared ]{
		\includegraphics[width = 0.4 \textwidth]{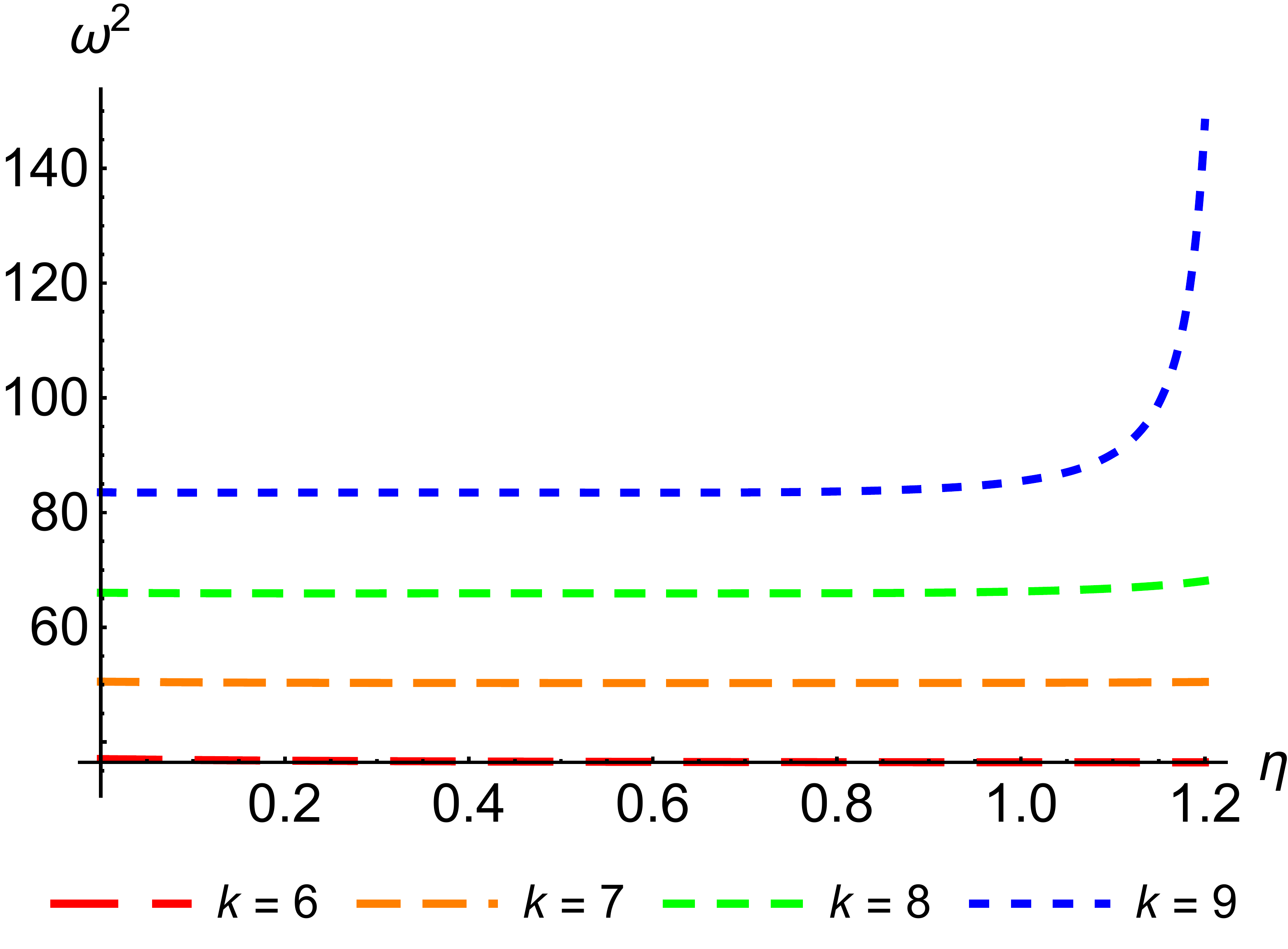}
		}
	\subfigure[ particle number ]{
		\includegraphics[width = 0.4 \textwidth]{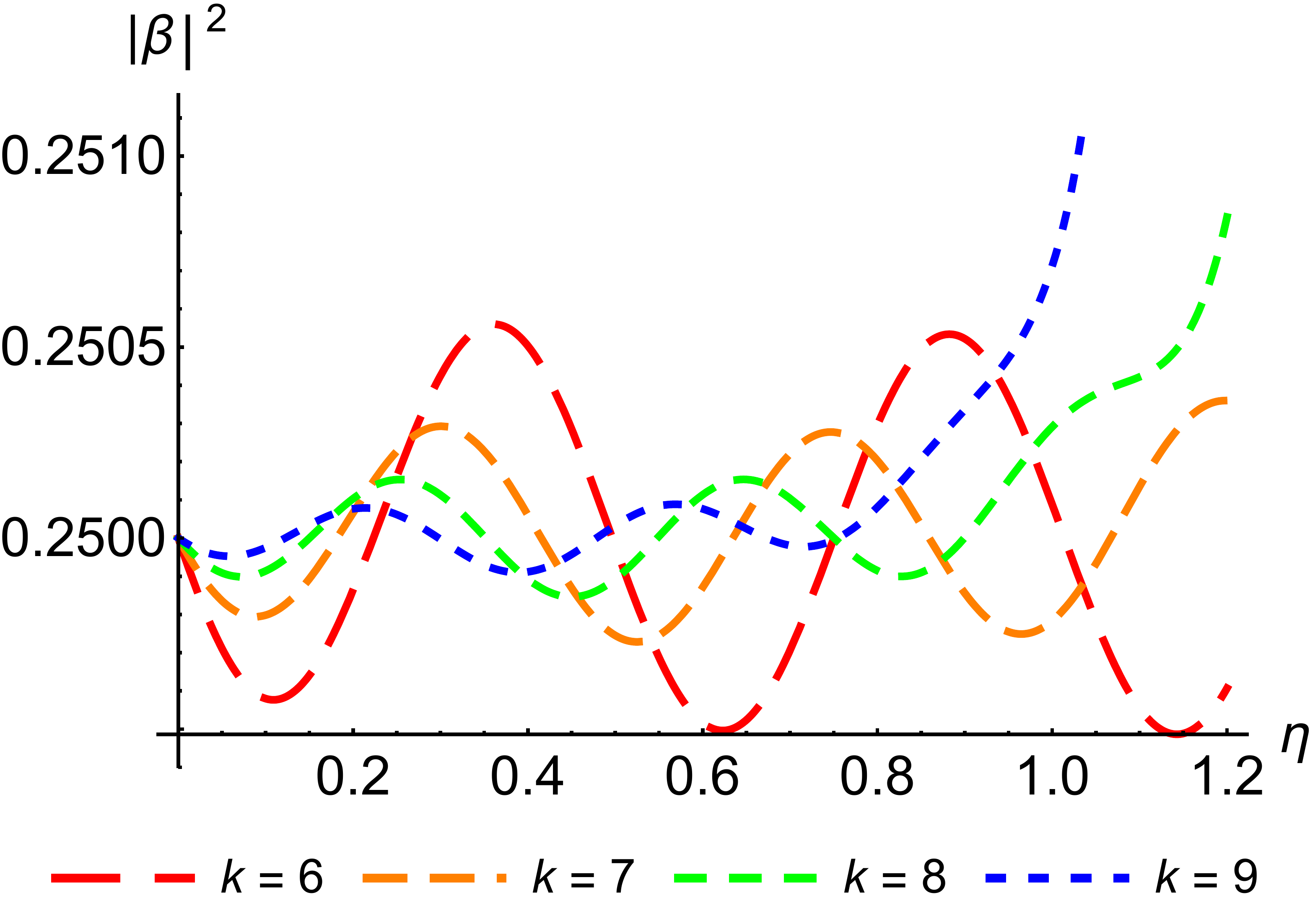}
		}
	\subfigure[ scalar field's energy density ]{
		\includegraphics[width = 0.4 \textwidth]{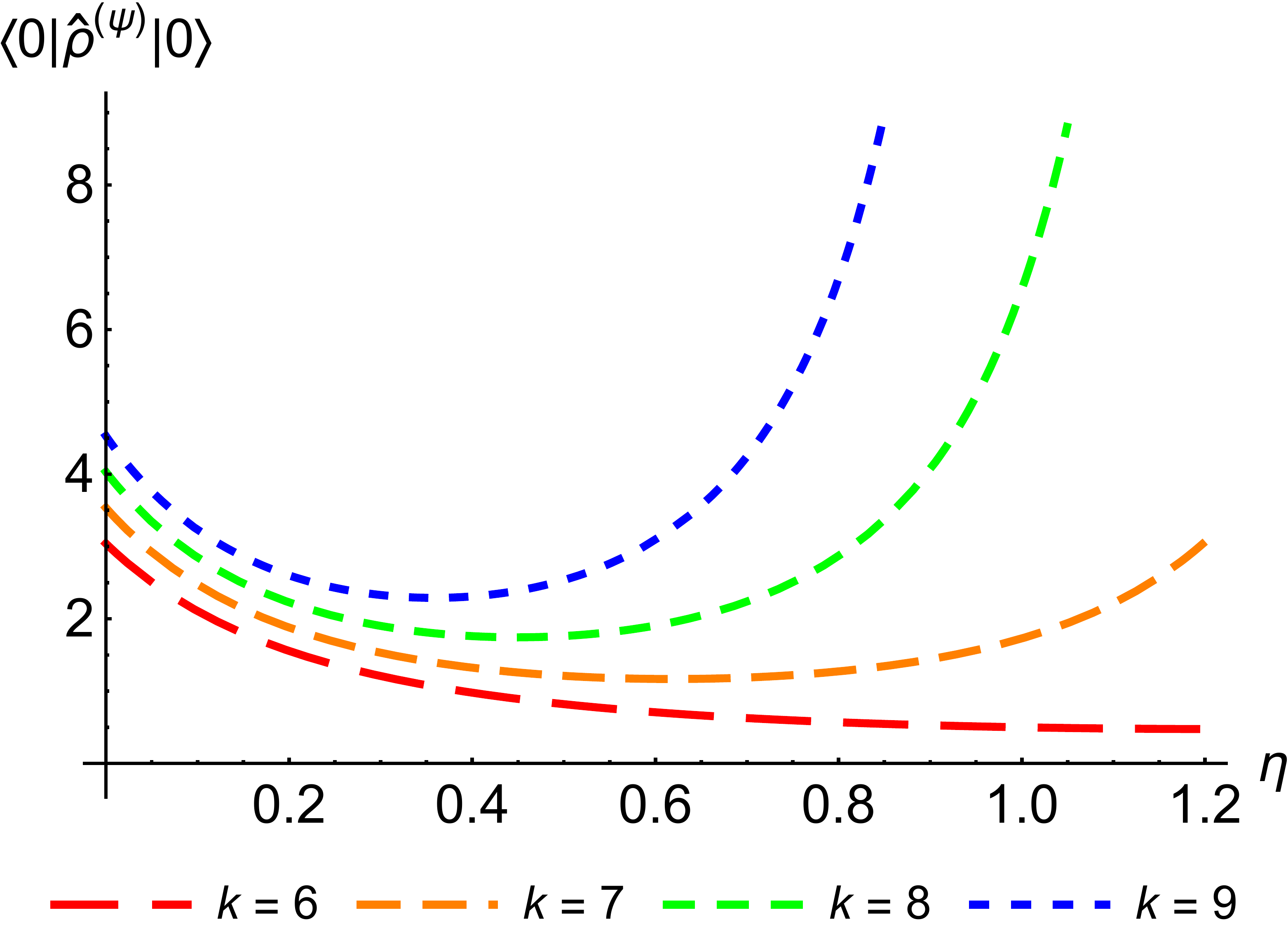}
		}
\caption{Plots of (a) scale factor, $a$, (b) Hubble parameter, $H = a'/a^2$, (c) QHO Hamiltonian, $\langle 0 | \hat{H} | 0 \rangle$, (d) frequency squared, $\omega^2$, (e) particle number, $| \beta |^2$, and (f) scalar field's energy density, $\langle 0 | \hat{\rho}^{\left(\psi\right)} | 0 \rangle$, obtained using the CQC for $k = 6, 7, 8, 9$ with $a_0 = 1$ and $h = 1$.}
\label{fig:short}
\end{figure*}
Figures \ref{fig:short}(a) and (b) for the scale factor and the Hubble parameter, respectively, confirm that the spacetime collapses due to the backreaction of short wavelength-modes. For all cases in figure \ref{fig:short}, however, the vacuum energy is negative and so this might be the reason why the integration leads to an eventually collapsing spacetime background. The discussion of the nearly critical wavelength-modes, on the other hand, have shown that even for $\rho > 0$ it can be inevitable to end up with a collapsing spacetime background because of the backreaction of the quantum field \footnote{
The precise solution to this apparent puzzle will be discussed in a different paper investigating the initial conditions of inflation in light of the CQC.
}. Figures \ref{fig:short}(c) and (d), moreover, show that short wavelength-modes are supported by a roughly constant Hamiltonian and frequency and are therefore more stable compared to the long wavelength and nearly critical wavelength counterparts. At the same time, the scalar field's energy density (figure \ref{fig:short}(f)) increases notably because of the collapse during the late stages of the evolution. This is accompanied by a production of modes (figure \ref{fig:short}(e)). It must also be mentioned that the larger $k$ is, or rather the shorter $\lambda < \lambda_\text{min}$ is, the earlier the spacetime collapses due to backreaction.

\subsubsection{An analytical solution for short-wavelength modes}
\label{subsubsec:analytical_solution}

The assumption of constant frequency, which is supported by the results for the short wavelength-modes (figure \ref{fig:short}(d)), can be used to obtain an approximate analytical solution and moreover deduce the existence of the threshold. This calculation can be done starting with Eq. \eqref{eq:omega_k_a} and taking the left hand side to be a constant that is completely determined by Eqs. \eqref{eq:omega_0_2} and \eqref{eq:rho_k_vac_0}. The vacuum energy can be determined by using the Friedmann constraint (Eq. \eqref{eq:fe1_conf}). This leaves us with a second-order linear ordinary differential equation for the scale factor, $a'' + \left( \omega_0^2 - k^2 \right) a = 0$, which incidentally is the harmonic oscillator's equation of motion itself. By using the same initial conditions as with the semiclassical solution (Eq. \eqref{eq:a_ds_classical}), we obtain the approximate short wavelength-mode solution
\begin{equation}
\label{eq:a_short_wavelength}
\begin{split}
a\left( \eta \right) = 
a_0 \cosh \left(\eta f\left( k \right) \right) + \dfrac{ a_0^2 h \sinh \left( \eta f\left( k \right) \right)}{ f \left( k \right) }
\end{split}
\end{equation}
where
\begin{equation}
f \left( k \right) = \dfrac{ \sqrt{8 a_0^4 h^2 \sigma -a_0^2 h^2-k^2-\sigma ^2}}{2 a_0 \sqrt{\sigma }} .
\end{equation}
This analytical expression (recall that $\sinh \left( x \right), \cosh \left( x \right) \sim e^{x} , x \rightarrow \infty$) shows that a de Sitter state, a.k.a., an inflationary era, would survive at late times even with backreaction provided that the quantity $f\left( k \right)$ is real or rather that 
\begin{equation}
8 a_0^4 h^2 \sigma -a_0^2 h^2-k^2-\sigma ^2 > 0 .
\end{equation}
Fixing $\sigma = \omega_0 \left( k \right)$, this therefore predicts a cutoff wavenumber, $k = k_\text{max}$, defined by 
\begin{equation}
k_{\text{max}}^2 = 8 a_0^4 h^2 \omega_0 \left( k_{\text{max}} \right) -a_0^2 h^2 - \omega_0 ^2 \left( k_{\text{max}} \right) ,
\end{equation}
or rather a threshold wavelength, $\lambda_\text{min} = 2\pi/k_{\text{max}}$, such that modes with $k > k_{\text{max}}$ would destabilize the background to collapse. This approximation understandably predicts $k_\text{max} \approx 3.87$ instead of $k_\text{max} \approx 5.77$ because the accuracy of this approximation is expected to improve only for large $k$, where the assumption of a constant frequency holds to a certain degree. Nonetheless, it supports the conclusion that there should be a threshold Planckian wavelength. 
\begin{figure}[h!]
\center
	\subfigure[ $k = 6$ ]{
		\includegraphics[width = 0.4 \textwidth]{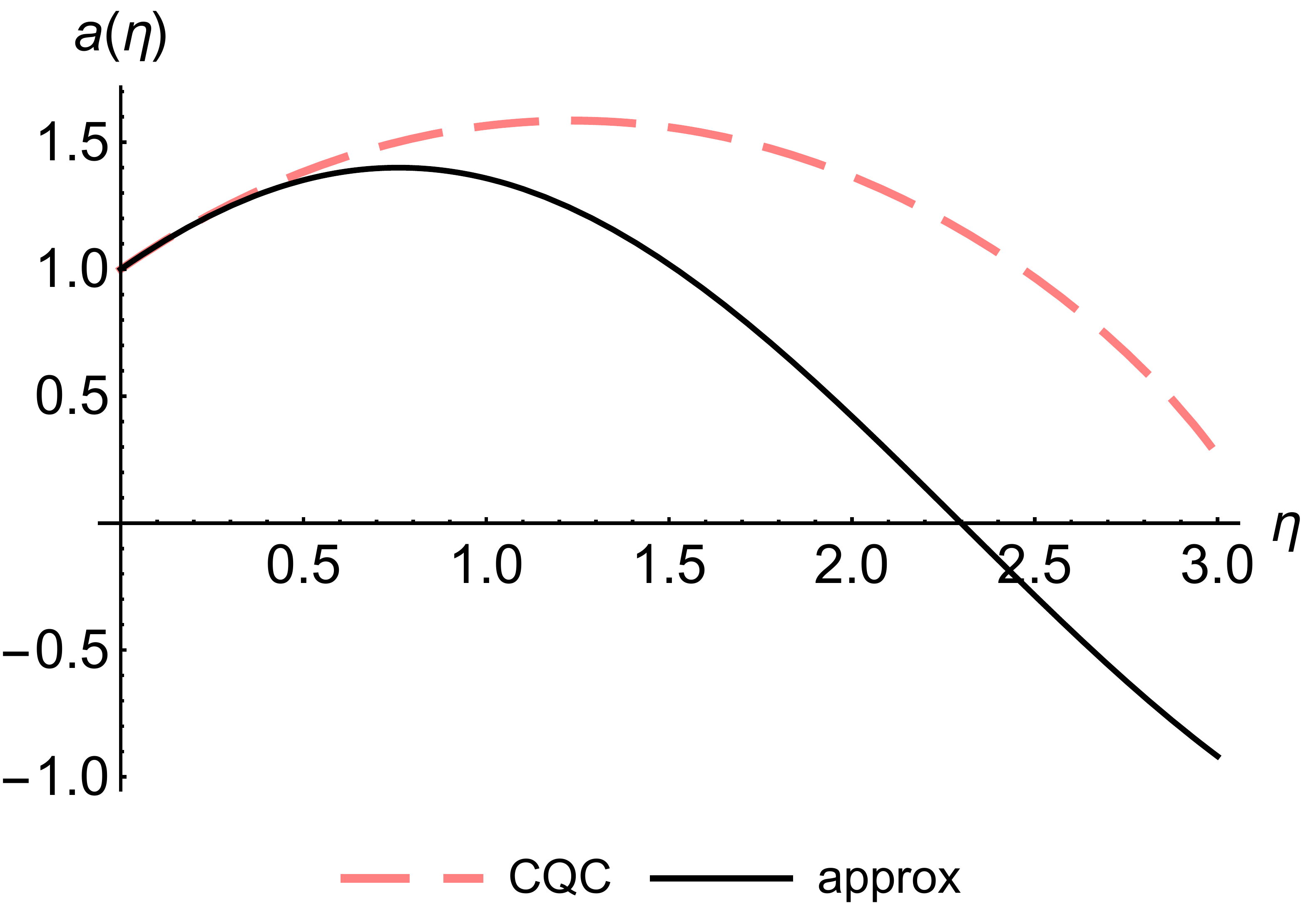}
		}
	\subfigure[ $k = 8$ ]{
		\includegraphics[width = 0.4 \textwidth]{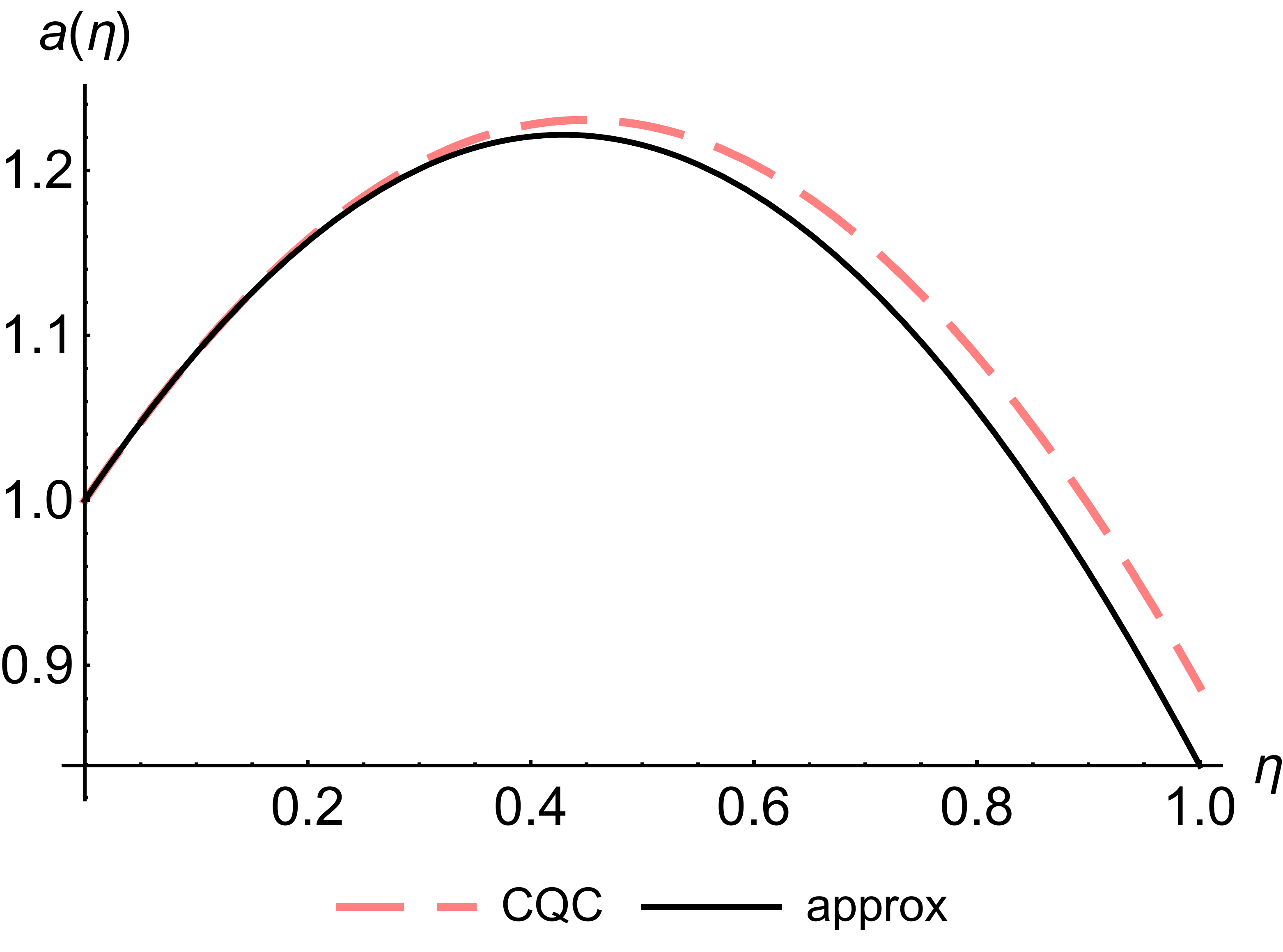}
		}
\caption{Scale factor obtained using Eq. \eqref{eq:a_short_wavelength} and the CQC for $k = 6$ and $k = 8$ with $a_0 = 1$ and $h = 1$.}
\label{fig:approx}
\end{figure}
Plots of the scale factor given by Eq. \eqref{eq:a_short_wavelength} and the CQC are shown in figure \ref{fig:approx} for $k = 6$ and $k = 8$. This confirms that the approximation (Eq. \eqref{eq:a_short_wavelength}) only starts to become better for larger $k$, which is to be expected based on the validity of the constant frequency assumption.

\section{CQC versus perturbative-iterative method}
\label{sec:cqc_iterative_method}

Prior to the CQC, a common way of obtaining semiclassical gravity effects is by solving for the Bogoliubov coefficients provided a fixed classical background. The solution to this, i.e., the Bogoliubov coefficients, could be used to obtain the leading order (LO) backreaction effect to the background which henceforth could be used to obtain the modes at next-to-leading order (NLO), and so on. This is the perturbative-iterative method.

It was shown that the CQC is the large iteration limit of the perturative-iterative method in the context of the rolling ball-QHO toy problem \cite{cqc_backreaction}. The same assertion was also argued to be the case for CQC's field theory counterpart \cite{cqc_fields}. In this section, we show that this is also the case for the Einstein-Klein-Gordon system. To obtain the perturbative-iterative results, we start with the de Sitter background given by Eq. \eqref{eq:a_ds_classical} and from this solve for $\left( \xi , \chi \right)$ using Eqs. \eqref{eq:xi_cho_eom} and \eqref{eq:chi_cho_eom}. We stress that this is the familiar elementary step of solving for the Bogoliubov coefficients, provided a fixed background, but expressed more conveniently in terms of the CHOs $\xi$ and $\chi$. Computing the mode's stress-energy tensor using $\xi$ and $\chi$, one obtains the leading order (LO) backreaction effect by solving the Friedmann equations (Eqs. \eqref{eq:friedmann_eq} and \eqref{eq:hubble_eq}) for the scale factor $a$. This scale factor can then again be used to obtain $\left( \xi, \chi \right)$ which will determine a better approximation to $a$ at next-to-leading order (NLO).
This procedure can be continued for any practical order in iteration to obtain a better approximation to the exact solution.

\begin{figure*}[h!]
\center
	\subfigure[ $k = 2$ ]{
		\includegraphics[width = 0.4 \textwidth]{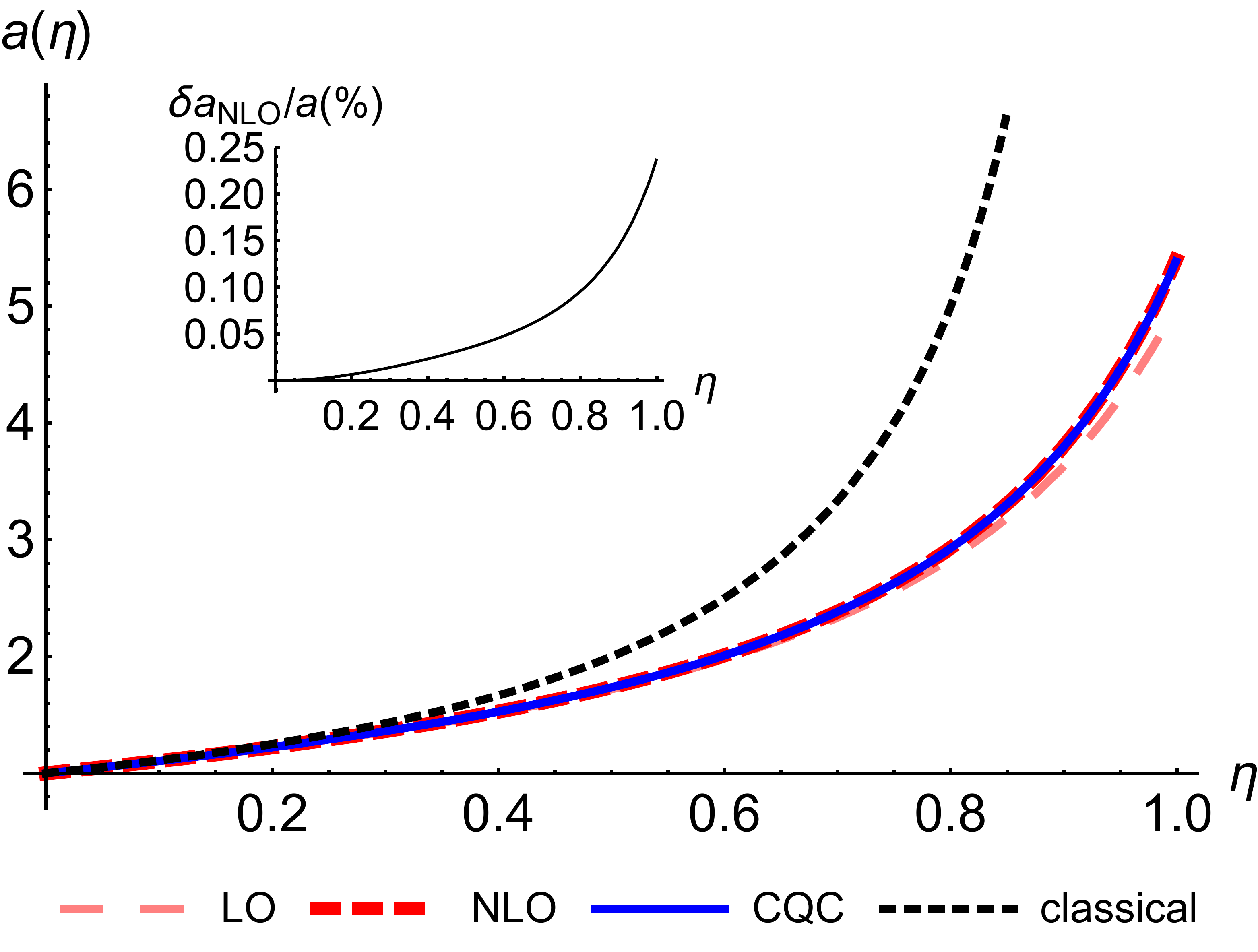}
		}
	\subfigure[ $k = 4$ ]{
		\includegraphics[width = 0.4 \textwidth]{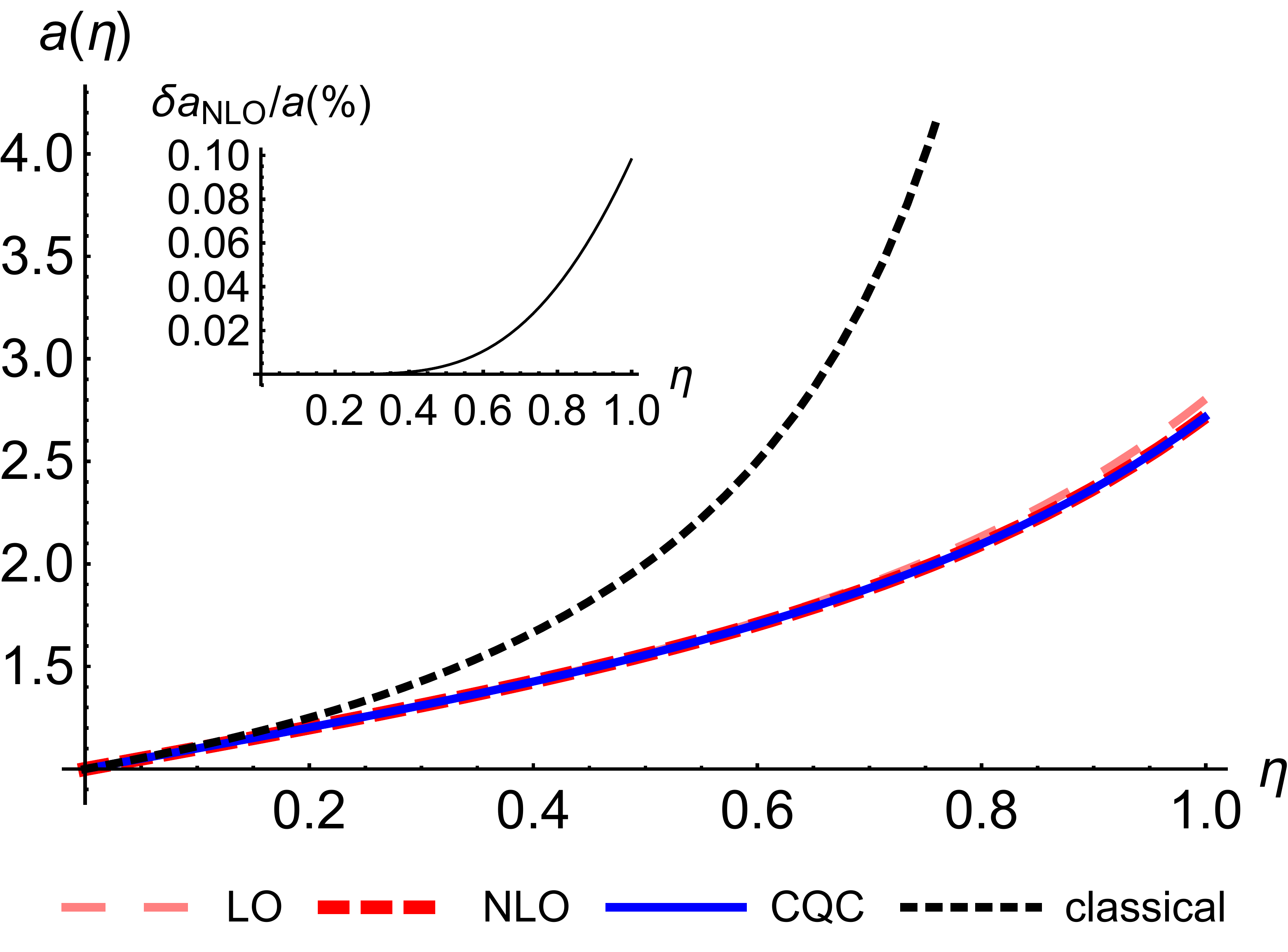}
		}
	\subfigure[ $k = 6$ ]{
		\includegraphics[width = 0.4 \textwidth]{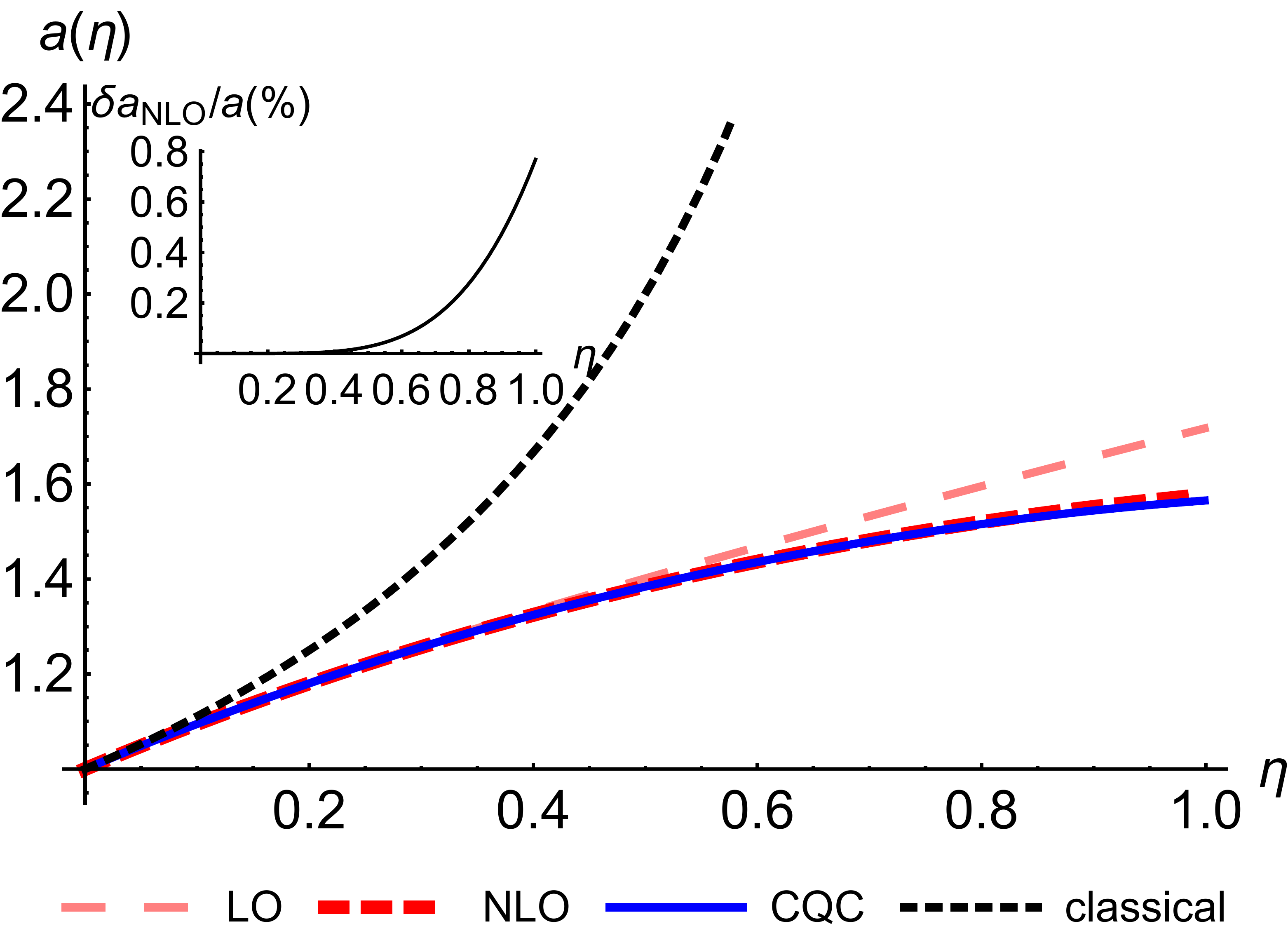}
		}
	\subfigure[ $k = 8$ ]{
		\includegraphics[width = 0.4 \textwidth]{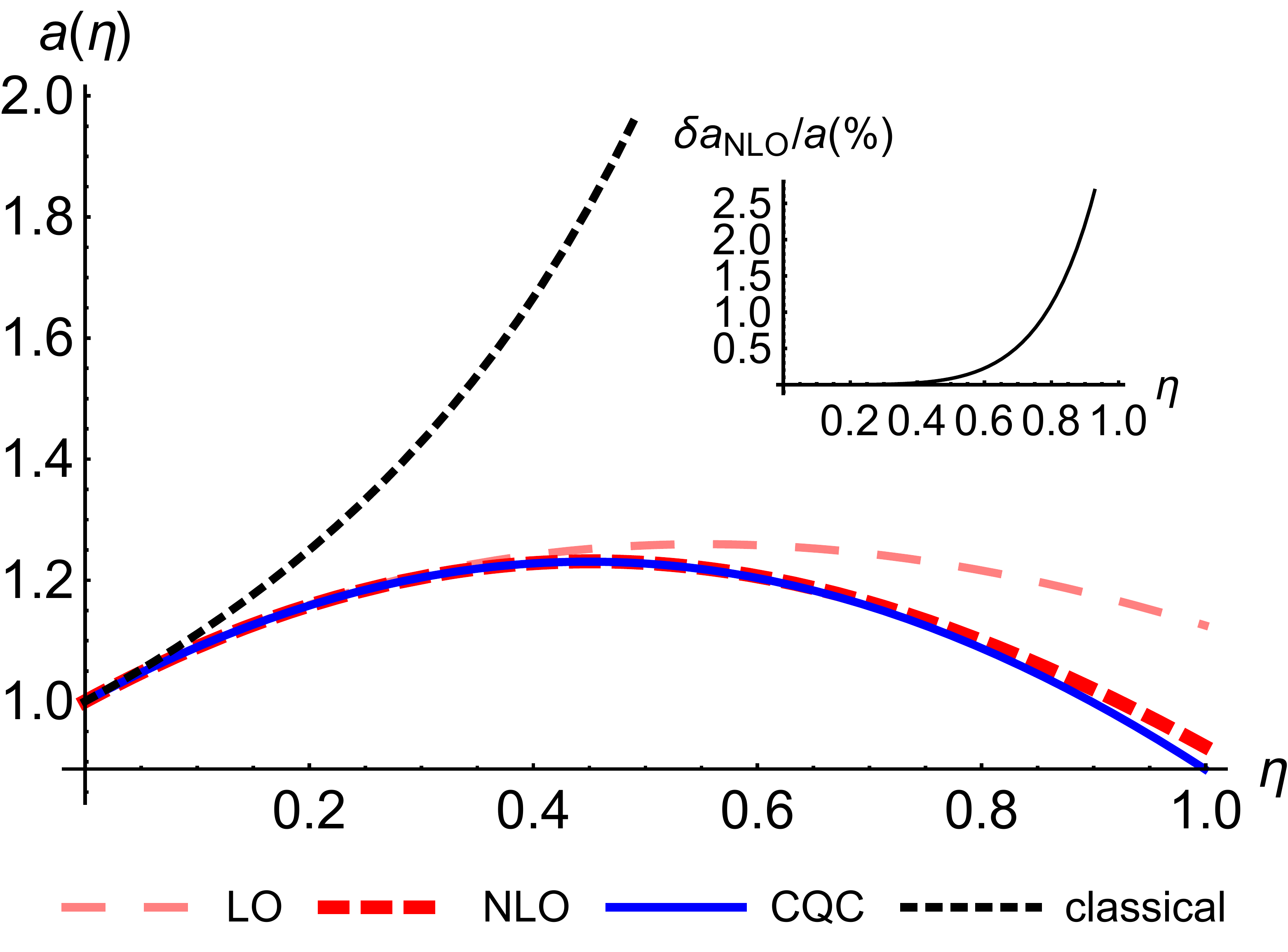}
		}
\caption{Scale factor, $a$, obtained using the CQC and the leading order (LO) and next-to-leading order (NLO) of the perturbative-iterative method for $k = 2, 4, 6, 8$ with $a_0 = 1$ and $h = 1$. Inset plot shows the relative deviation of the NLO result from the CQC, $\delta a_{\text{NLO}}/a = | \left( a_\text{NLO} - a_\text{CQC} \right) / a_\text{CQC} | \times 100 \%$, as the results are impossible to visually distinguish for long wavelength-modes.}
\label{fig:a_NLO}
\end{figure*}

\begin{figure*}[h!]
\center
	\subfigure[ $k = 2$ ]{
		\includegraphics[width = 0.4 \textwidth]{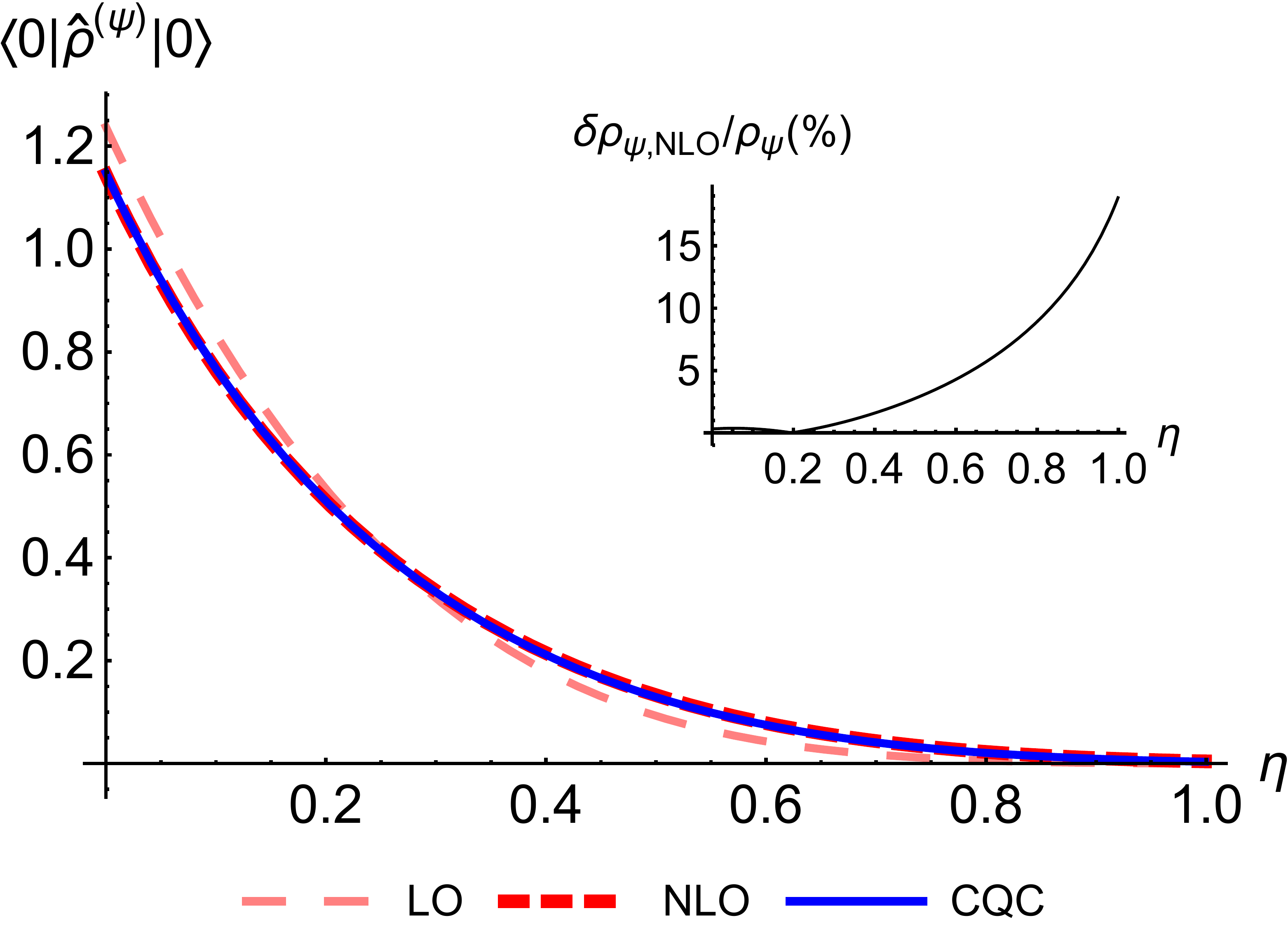}
		}
	\subfigure[ $k = 4$ ]{
		\includegraphics[width = 0.4 \textwidth]{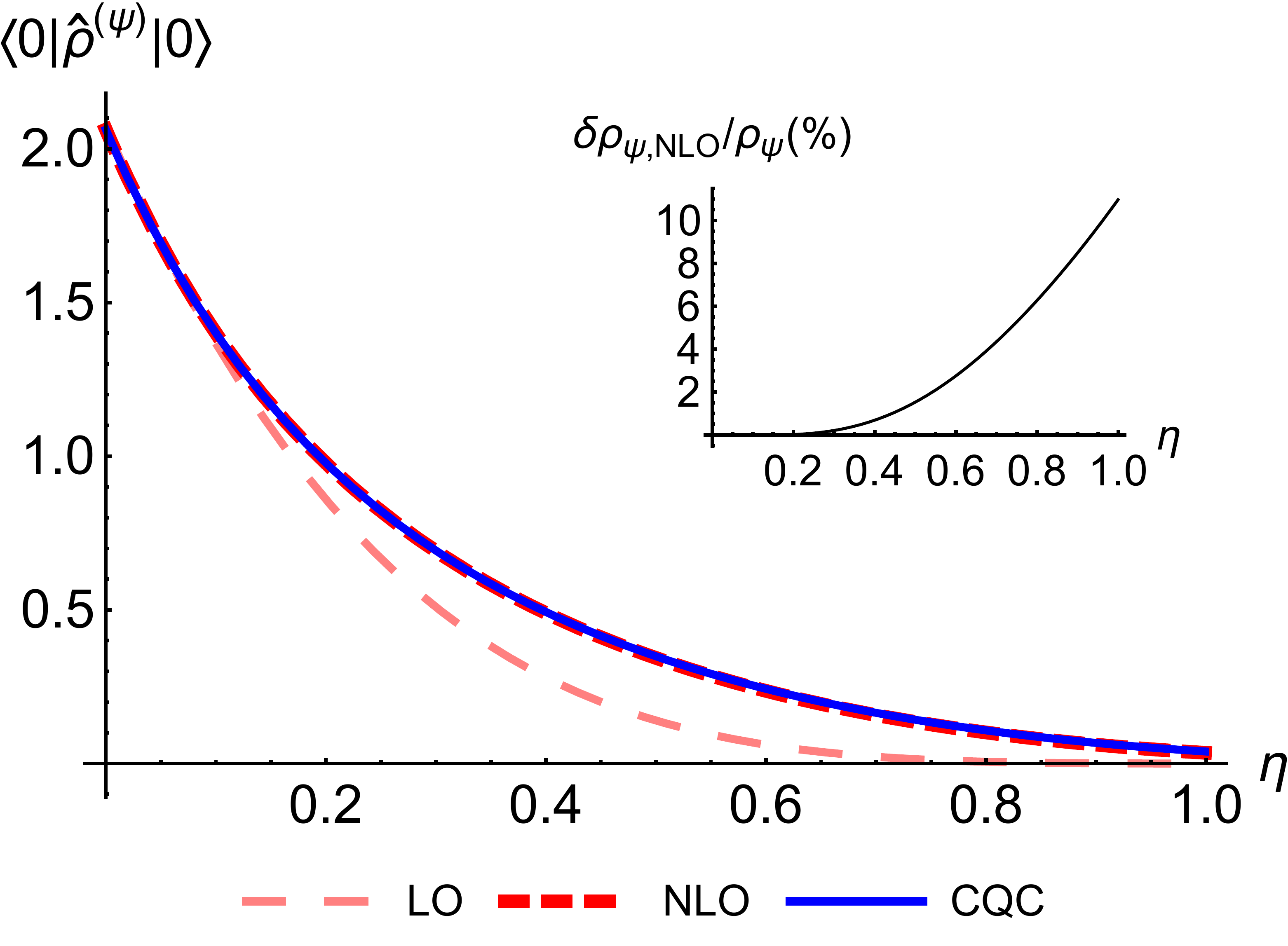}
		}
	\subfigure[ $k = 6$ ]{
		\includegraphics[width = 0.4 \textwidth]{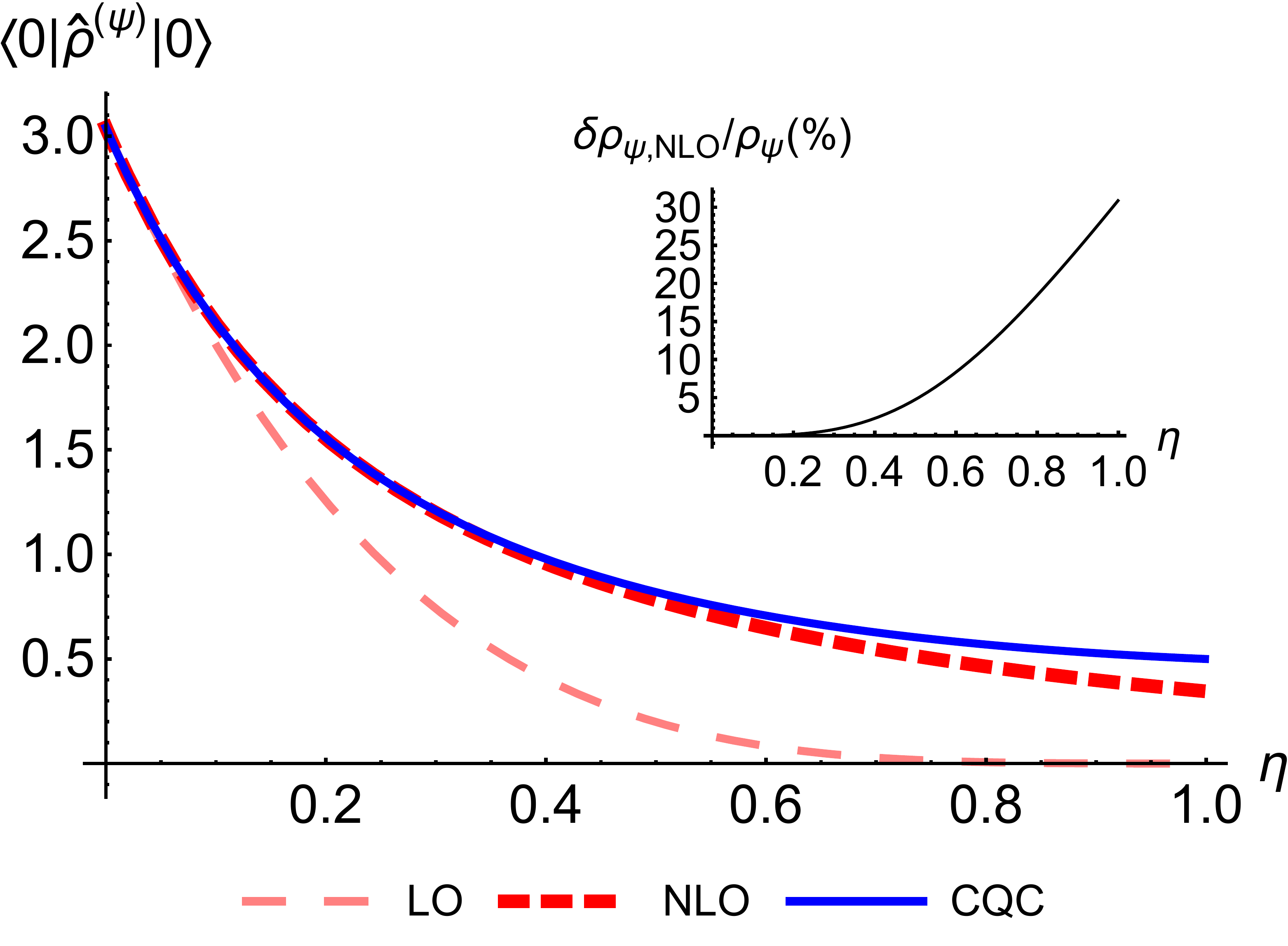}
		}
	\subfigure[ $k = 8$ ]{
		\includegraphics[width = 0.4 \textwidth]{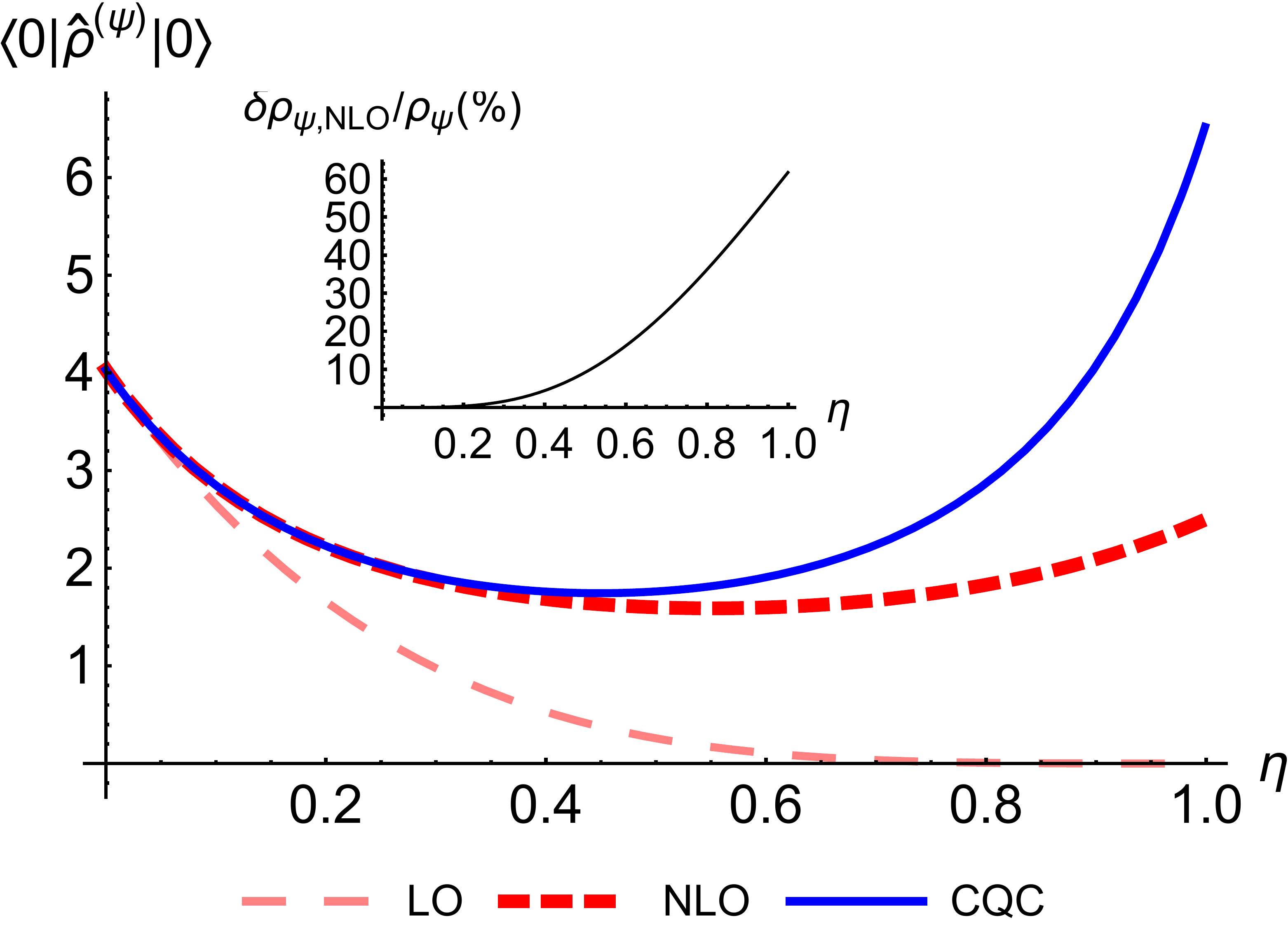}
		}
\caption{Scalar field's energy density, $\rho_\psi = \langle 0 | \hat{\rho}^{\left(\psi\right)} | 0 \rangle$, obtained using the CQC and the leading order (LO) and next-to-leading order (NLO) of the perturbative-iterative method for $k = 2, 4, 6, 8$ with $a_0 = 1$ and $h = 1$. Inset plot shows the relative deviation of the NLO result from the CQC, $\delta \rho_{\psi, \text{NLO}}/ \rho_\psi = | \left( \rho_{\psi,\text{NLO}} - \rho_{\psi, \text{CQC}} \right) / \rho_{\psi,\text{CQC}} | \times 100 \%$, as the results are impossible to visually distinguish for long wavelength-modes.}
\label{fig:rho_NLO}
\end{figure*}

Figures \ref{fig:a_NLO} and \ref{fig:rho_NLO} show the LO and NLO results for the scale factor, $a$, and the scalar field's energy density, $\langle 0 | \hat{\rho}^{(\psi)} | 0 \rangle$, together with the CQC.
All of the cases considered indeed strongly confirm that perturbative-iterative method approaches the CQC. Remarkably, even at NLO ($N = 2$ iterations) only, the agreement between the scale factor and the CQC counterpart can be considered to be satisfactory, most especially for long wavelength modes where it becomes impractical to visually distinguish the NLO result and CQC. In particular, the relative deviation of the NLO from the CQC for the scale factor peaks only at $0.25\%$, $0.10\%$, $0.80\%$, and $2.5\%$ for $k = 2, 4, 6, 8$, respectively. The subpercent error level explains why the plots of the NLO and CQC cannot be visually distinguished, hence, the need to show the inset plots of relative error in figure \ref{fig:a_NLO}. A similar result can be observed between the energy density, $\langle 0 | \hat{\rho}^{(\psi)} | 0 \rangle$, obtained at NLO and the CQC, although with a notably larger deviation as compared with the scale factor results. This, nonetheless, continues to support the assertion that the CQC is the large iteration limit of perturbative-iterative method as the NLO can be observed to approach the CQC in all cases. The visual indistinguishability of the NLO and CQC results, especially for longer wavelengths, demands again the need to show the inset of relative error in figure \ref{fig:rho_NLO}. At NLO, thus, the peak relative error from the CQC is at $15\%$, $10\%$, $30\%$, and $60\%$ for $k = 2, 4, 6, 8$, respectively. The relative deviation clearly increases for shorter wavelengths (larger $k$) where backreaction becomes more significant. This calls for larger iteration orders in the perturbative-iterative method for short wavelength-modes.

\begin{figure}[h!]
\center
	\subfigure[ $\delta a_{N = 10}/a = | \left( a_{N = 10} - a_\text{CQC} \right) / a_\text{CQC} | $ ]{
		\includegraphics[width = 0.4 \textwidth]{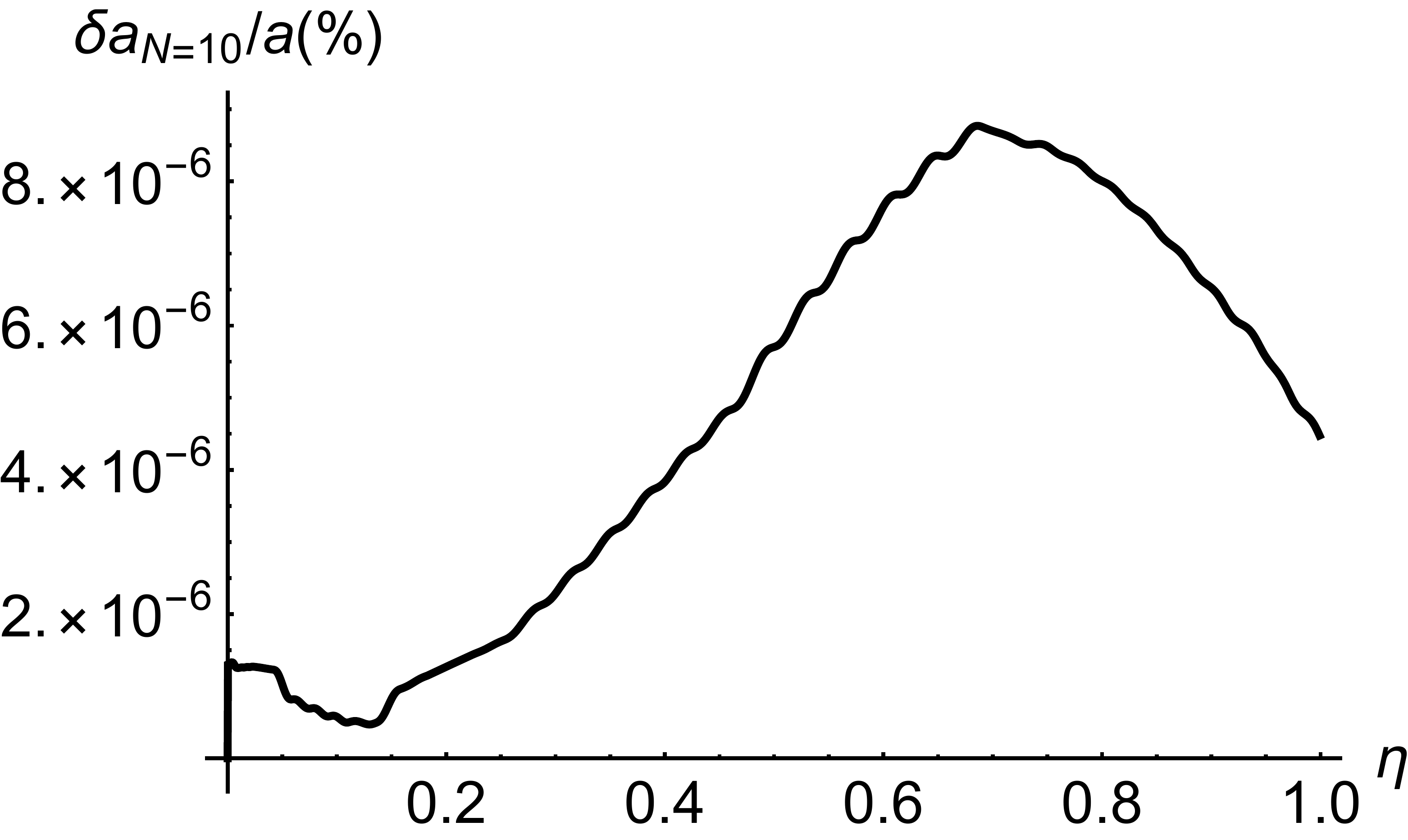}
		}
	\subfigure[ $\delta \rho_{\psi, N = 10}/ \rho_\psi = | \left( \rho_{\psi,N = 10} - \rho_{\psi, \text{CQC}} \right) / \rho_{\psi,\text{CQC}} |$ ]{
		\includegraphics[width = 0.4 \textwidth]{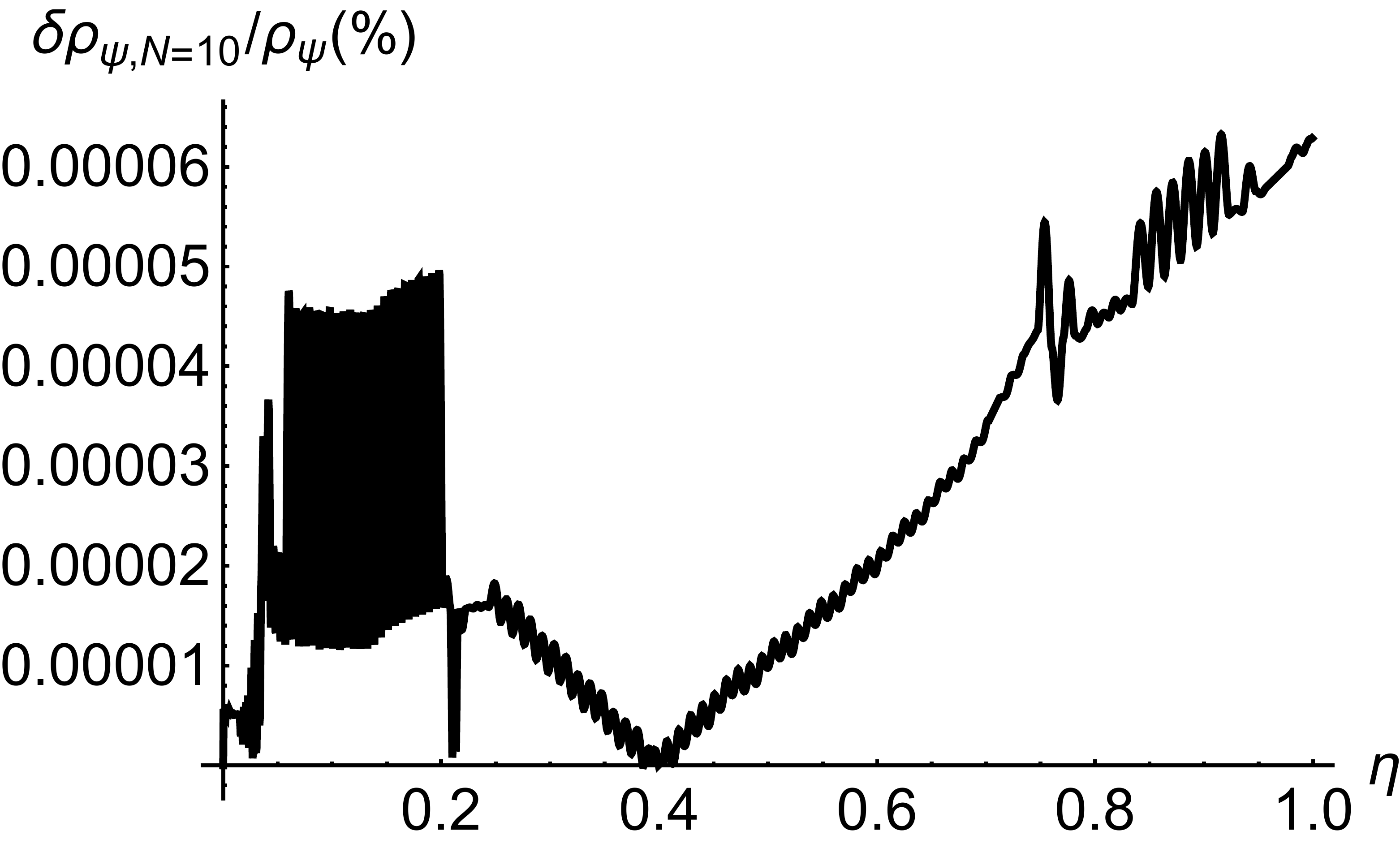}
		}
\caption{Relative error between the tenth-order ($N = 10$) iterative-perturbative method and CQC for the scale factor, $a$, and the scalar field's energy density, $\langle 0 | \hat{\rho}^{\left( \psi \right)} | 0 \rangle$ for $k = 8$ with $a_0 = 1$ and $h = 1$.}
\label{fig:flex}
\end{figure}

Building on the NLO, we demonstrate that the CQC is the undeniable limit of the perturbative-iterative method by comparing the tenth order ($N = 10$) iteration and the CQC for the scale factor, $a$, and the scalar field's energy density, $\langle 0 | \hat{\rho}^{(\psi)} | 0 \rangle$. This result is shown in figure \ref{fig:flex} for $k = 8$ which was previously where the NLO can be regarded as an awkward approximation.
The subpercent agreement ($\delta a, \delta \rho_\psi \sim 10^{-5}\%$) shown in figure \ref{fig:flex} strongly supports the CQC as the limit of the perturbative-iterative method. 

\section{Discussion}
\label{sec:discussion}

We have reviewed the CQC in detail (Section \ref{sec:cqc}) and, for the first time, successfully applied it to a quantum field that is coupled to a cosmological background (Section \ref{sec:backreaction_ds}). 
In addition, we have shown that the CQC is the undeniable limit of the perturbative-iterative method in semiclassical gravity (Section \ref{sec:cqc_iterative_method}), supporting previous assertions of this statement in the rolling ball-QHO system \cite{cqc_backreaction} and CQC's field theory extension \cite{cqc_fields}.

This is also a good place to remind that we have not dealt with renormalization thus far and have considered only the individual dynamics of each of the modes of the scalar field $\hat{\phi}(x)$ in an expanding spacetime background. Each of these modes were then taken to be initially in their lowest energy state which is a superposition of the growing and decaying modes familiar in inflationary studies focusing on the transition of quantum mechanical modes to classical degrees of freedom. The backreaction effects presented in this paper must therefore be carefully interpreted in a per mode basis.

The application of the results of the paper are physically restricted to the domain of the coupled Einstein-(massless) Klein-Gordon system (Eqs. \eqref{eq:wave_equation} and \eqref{eq:einstein_equation}). The early Universe for all we know might be filled with more than just the inflaton and it is of interest to extend this work later to study the backreaction of spin-$1/2$ fields and massive vector fields to the spacetime. Nonetheless, the Einstein-(massless) Klein-Gordon system may also be able to draw information on the backreaction of massless higher-spin fields, e.g., each of the two polarization modes of gravitational waves generated during the inflationary era satisfies $u'' + ( k^2 - a''/a ) u = 0$, which is the same equation as the massless scalar field (Eq. \eqref{eq:kg_conformal_k}). The case of a massive scalar field, with a bare mass $m$, could also be very simply accommodated in the model by transforming the wavenumber $k$ as $k \rightarrow \sqrt{k^2 + m^2}$. Through this simple transformation it can be realized that the existence of a short wavelength threshold translate to the existence of a heavy mass cutoff that would destabilize the inflationary era.

There is one more feature that dissociates this work from the realistic scenario but one which we heavily clinged on to for great simplification. We have assumed from Eq. \eqref{eq:rho_psi_general} onwards that the quantum field lives only in one of its Fourier modes defined by Eq. \eqref{eq:fourier}.
However, the more realistic picture will always be one which involves a nontrivial superposition of Fourier modes, say for instance, a Gaussian distribusion centered at $k = K$ with a width $\sigma_K$. To proceed in this way demands an extension of the CQC to a system with interacting QHOs each of which is also coupled to the same spacetime background. Alternatively, the construction of a spatial lattice could be established as in the field theory-extension of the CQC \cite{cqc_fields, cqc_rolling_scalar, cqc_breathers}. Either way is more computationally exhausting but is irrefutably a relevant future work. 

A potential application of the CQC is on the preheating era where the quantum nature of the modes play a significant role \cite{inflation_preheating_armendariz}. The CQC can be used to improve the existing numerical recipes (see for instance Refs. \cite{latticeeasy, pspectre}) by taking into consideration the quantum nature of the matter fields and could be used to study \textit{nonperturbatively} the backreaction of quantum fields to both the inflaton and the metric.

The backreaction can also be used to gain insight on the primordial power spectrum. The modes of relevance to this are those which start very deep inside the horizon but then eventually cross the horizon and become a constant. During the very early inflationary era, the perturbations of the metric can be shown to be subdominant when compared to that of the inflaton \cite{weinberg}, and so the model used in this paper can be considered to be an approximate working model. Accounting for backreaction, the condition for horizon crossing of a comoving $k$-mode will therefore change from the naive $k / a( \eta_* ) = h$ to $ k/ a( \eta_* ) = H_{\text{eff}}( \eta_* )$ where $H_{\text{eff}} ( < h)$ is the renormalized Hubble parameter and $\eta_*$ is the conformal time at horizon crossing. The two-point correlation function of a quantum field, $\hat{ \phi }$, which gives rise to the primordial power spectrum, will then become proportional to $H_{\text{eff}}( \eta_* )^2$ rather than to $h^2$. The detailed analysis of this is left for future work.

\section{Conclusions}
\label{sec:conclusions}

We have applied the CQC, for the first time, to a system described by a quantum field that is driven by a cosmological background (Eqs. \eqref{eq:wave_equation} and \eqref{eq:einstein_equation}) in order to draw insigths on quantum mechanical backreaction during inflation. The results have shown that there exists a minimum Planckian wavelength that distinguishes between long wavelength-modes that would only renormalize the Hubble parameter and short wavelength-modes that would destabilize the inflationary era. We have also obtained an approximate analytical solution for the short-wavelength modes which support the existence of such a threshold wavelength. The results of the CQC and the standard perturbative-iterative method was also compared. This lead to a strong support to the assertion that the CQC is the inevitable limit of the perturbative-iterative method.

The model used in this paper can be understood to be a working theory of the very early Universe when the modes are deep inside the Hubble horizon and the metric perturbations can be discarded compared to the \textit{inflaton} perturbation. Nonetheless, it can be improved in many ways to be a more faithful representation of the early Universe. As a start, it would be desirable to extend this work to involve more than just a single mode backreacting to the metric. Also, we have relied so far on a fluid that is vacuum energy-like, to initially support the inflationary era. However, it is known that primordial inflation is more convincingly supported by a quantized slowly-rolling scalar field (a.k.a., the inflaton) in a potential that coexists with the spacetime and perhaps with some other fields in the early Universe. It is interesting to check out whether the addition of other fields could usher the inflationary era gracefully to the radiation era. The work must therefore be later on extended to viable inflationary theories (see for instance Refs. \cite{inflation_review_brandenberger, inflation_review_martin}) and compute the resulting primordial power spectrum and adiabatic invariants when the modes exit the horizon. Lastly, backreaction is a nontrivial effect even outside the context of inflation. The application of the CQC to simpler physical systems will therefore continue to be interesting and encouraged in future works.


\section*{Acknowledgements}
The author would like to thank two anonymous referees for constructive feedback which significantly improved the quality of the paper.

\appendix

\section{Planck units}
\label{sec:planck_units}

Planck units can be achieved by setting the fundamental physical constants $c$, $G$, and $\hbar$ to unity. Doing so keeps the math elegant and all of the physical quanties would be expressed in terms of their natural Planckian values, e.g., Planck length, time, and mass. 

To illustrate this, we set $c = 8 \pi G = \hbar = 1$. The units of length, time, and mass would then be in terms of their Plankian values,
\begin{eqnarray}
L_P &=& \sqrt{ \dfrac{ \tilde{ G } \hbar }{c^3} } \sim 10^{-35} \ \text{m} \\
T_P &=& \sqrt{ \dfrac{ \tilde{ G } \hbar }{c^5} } \sim 10^{-43} \ \text{s} \\
M_P &=& \sqrt{ \dfrac{ c \hbar }{ \tilde{ G } } } \sim 10^{-9} \ \text{kg}
\end{eqnarray}
where $\tilde{ G } = 8 \pi G$. The Planck energy can also be easily computed as $E_P = M_P c^2 \sim 10^{19} \ \text{GeV}$. It is straightforward to navigate to and from Planck units by simply multiplying factors of the above fundamental units of nature.



\end{document}